\documentclass[12pt,preprint]{aastex}
\begin{document}

\slugcomment{Accepted for publication in the Astrophysical Journal}

\title{A Uniform Analysis of the Ly-$\alpha$ Forest at $z=0 - 5$: \\
V. The extragalactic ionizing background at low redshift}

\author{Jennifer Scott, Jill Bechtold, Miwa Morita}
\affil{Steward Observatory, University of Arizona, Tucson, AZ 85721, USA \\
e-mail: [jscott,jbechtold,mmorita]@as.arizona.edu}

\author{Adam Dobrzycki}
\affil{Harvard-Smithsonian Center for Astrophysics, 60 Garden Street, \\
Cambridge, MA 02138, USA \\
e-mail: adobrzycki@cfa.harvard.edu}

\author{Varsha P. Kulkarni}
\affil{University of South Carolina, Department of Physics and Astronomy,  \\
Columbia, SC  29208, USA  \\
e-mail: varsha@mail.psc.sc.edu}

\begin{abstract}

In Paper III of our series $``$A Uniform Analysis of the 
Ly-$\alpha$ forest at $z=0 - 5$", we presented
a set of 270 quasar spectra from the archives of the
Faint Object Spectrograph (FOS) on the Hubble Space Telescope (HST). 
A total of 151 of these spectra, yielding 906 lines, are suitable for using the 
proximity effect signature to measure $J(\nu_{0})$, the mean intensity of the 
hydrogen-ionizing background radiation field, at low redshift.
Using a maximum likelihood technique and 
the best estimates possible for each QSO's Lyman limit flux and
systemic redshift,  we find
$J(\nu_{0})= 7.6^{+9.4}_{-3.0} \times 10^{-23}$ ergs s$^{-1}$ cm$^{-2}$ Hz$^{-1}$ sr$^{-1}$
at $0.03 < z < 1.67$.  This is in good agreement with the mean intensity expected
from models of the background which incorporate only the known quasar population.
When the
sample is divided into two subsamples, consisting of lines with
$z < 1$ and $z > 1$,  the values of $J(\nu_{0})$ found are
6.5$^{+38.}_{-1.6} \times 10^{-23}$ ergs s$^{-1}$ cm$^{-2}$ Hz$^{-1}$ sr$^{-1}$,
and
1.0$^{+3.8}_{-0.2} \times 10^{-22}$ ergs s$^{-1}$ cm$^{-2}$ Hz$^{-1}$ sr$^{-1}$,
respectively, indicating that the mean intensity of the
background is evolving over the redshift range of this data set.
Relaxing the assumption that the spectral shapes of the sample spectra and the
background are identical, the best fit HI photoionization rates are found to be
$6.7 \times 10^{-13}$ s$^{-1}$ for all redshifts, and $1.9 \times 10^{-13}$ s$^{-1}$
and $1.3 \times 10^{-12}$ s$^{-1}$ for $z < 1$ and $z > 1$, respectively.
The inclusion of blazars, associated absorbers, or
damped Ly-$\alpha$ absorbers, or the consideration of a $\Lambda$ CDM cosmology
rather than one in which $\Omega_{\Lambda}=0$ has no significant effect on the results. The 
result obtained using radio loud objects is not significantly different from
that found using radio quiet objects. 
Allowing for a variable equivalent width threshold gives a consistently
larger value of $J(\nu_{0})$ than the constant threshold treatment, though
this is found to be sensitive to the inclusion of a small number of weak lines
near the QSO emission redshifts.
This work confirms that the evolution of the number density
of Ly-$\alpha$ lines is driven by a decrease in the ionizing background
from $z \sim 2$ to $z \sim 0$ as well as by the formation of structure in the
intergalactic medium.   

\end{abstract}

\keywords{diffuse radiation--- intergalactic medium---quasars: absorption lines}

\section{Introduction}
\label{sec-intro}
The spectra of quasars show a $``$forest" of absorption lines blueward of
the Ly-$\alpha$ emission line (Lynds 1971, Sargent et al.\ 1980, Weymann, Carswell,
\& Smith 1981).
Observational and theoretical work in recent years has shown that most of this 
absorption can be attributed to neutral
hydrogen in galaxies and large-scale structure along the line of sight
(Cen et al.\ 1994, Lanzetta et al.\ 1995,1996, Zhang et al.\ 1995, 
Hernquist et al.\ 1996, 
Miralda-Escud\'{e} et al.\ 1996, Bi \& Davidsen 1997,
Chen et al.\ 1998, Theuns et al.\ 1998, Ortiz-Gil et al.\ 1999,  Impey, Petry, \& Flint 1999,
Dav\'{e} et al.\ 1999, Bryan et al.\ 1999).
In aggregate, QSO spectra
show an increasing line density with increasing redshift such that $dN/dz  \propto
(1.+z)^{\gamma}$ (Sargent et al.\ 1980, Weymann, Carswell, \& Smith 1981, Young et al.\ 1982,
Murdoch et al.\ 1986, Lu, Wolfe, \& Turnshek, 1991,
Bechtold 1994, Kim et al.\ 1997).
But the line density within an individual quasar spectrum
decreases with proximity to the Ly-$\alpha$ emission line (Weymann, Carswell, \& Smith 1981, 
Murdoch et al.\ 1986).
This is generally thought to be due to enhanced ionization of neutral hydrogen
in the vicinity of the quasar due to ionizing photons from the quasar itself.
This $``$proximity effect" can be used to measure the
mean intensity of the UV background, denoted $J(\nu_{0})$ (Carswell et al.\ 1987, Bajtlik, Duncan,
\& Ostriker 1988, hereafter BDO).

$J(\nu_{0})$ has been measured at $z > 1.7$
by a variety of authors 
(BDO, Lu, Wolfe, \& Turnshek 1991, Giallongo et al.\ 1993,1996,
Bechtold 1994, Williger et al.\ 1994,
Cristiani et al.\ 1995, Fern\'{a}ndez-Soto et al.\ 1995, Lu et al.\ 1996, Savaglio et al.\ 1997,
Cooke et al.\ 1997, Scott et al.\ 2000b).
The results, summarized in Paper II of this series (Scott et al.\ 2000b), are in general
agreement with the predictions of models of the UV background which 
integrate the contribution from
known population of quasars and include reprocessing effects in
an inhomogeneous intergalactic medium
(Haardt \& Madau 1996, hereafter HM96, Fardal et al.\ 1998).  In Paper II,
the mean intensity of the ionizing background was
found to be $7.0^{+3.4}_{-4.4} \times 10^{-22}$ 
ergs s$^{-1}$ cm$^{-2}$ Hz$^{-1}$ sr$^{-1}$ at $z \sim 3$.
The decline of the
quasar space density from $z \sim 2$ to the present is expected
to drive a corresponding decline in the intensity of the UV background.
Kulkarni \& Fall (1993, hereafter KF93) measured $J(\nu_{0}) \sim 6
\times 10^{-24}$ ergs s$^{-1}$ cm$^{-2}$ Hz$^{-1}$ sr$^{-1}$ at $z \sim 0.5$ from
a subset of the now complete HST Quasar Absorption Line Key Project
sample presented by Bahcall et al.\ (1993).

Much of this previous work has relied upon
the technique for measuring $J(\nu_{0})$ outlined by BDO.
This technique requires the entire sample of absorption lines  to
be binned according to the
ratio of the quasar flux at the physical position of the absorber to the background flux.  
This is done for several initial guesses of the background intensity; and the value
that gives the lowest $\chi^{2}$ between the binned data and the ionization model
is chosen as the best fit $J(\nu_{0})$.
However, this is not
the optimal technique to use at low redshift where absorption line
densities are low. 
KF93 developed
a maximum likelihood technique to address this issue and used it
in their measurement of $J(\nu_{0})$ at $z \sim 0.5$.
However, their measurement was based upon 
a sample of only 13 QSOs and less than 100 lines,
and has correspondingly large error bars.  In addition, the value these 
authors find is lower than the predictions of the models of Haardt \& Madau (1996),
though consistent within the uncertainties,
as shown in Figure 13 of Paper II and in Figure~\ref{fig:lowzcomp} of this paper.
Given the importance of the value of the HI ionization rate to the
hydrodynamical evolution of the low redshift universe,
performing this measurement
with a much larger line sample is worthwhile.

The low redshift hydrodynamic simulations of Theuns et al.\ (1998) and
Dav\'{e} et al.\ (1999) indicate that
the evolution of the ionizing background is the primary driver
behind the change of character of the Ly-$\alpha$ forest from high
redshift to low redshift, specifically, the break in the number
distribution of Ly-$\alpha$ lines at $z=1.7$ (Morris et al.\ 1991,
Bahcall et al.\ 1991, Weymann et al.\ 1998).
The growth of structure pulling gas from low density regions 
into high density regions also contributes to this and
other attributes of the evolution of the Ly-$\alpha$ forest.

Shull et al.\ (1999) estimate the local ionizing background
including contributions to the background
from starburst galaxies as well as Seyferts and QSOs.  Their models
include a treatment of the opacity of the low redshift Ly-$\alpha$ forest
using information drawn from recent observational work
(Weymann et al.\ 1998, Penton et al.\ 2000b).
They find that starbursts and AGN could contribute approximately equally
to the ionizing background at low redshift, each $\sim 1.0 \times 10^{-23}$
ergs s$^{-1}$ cm$^{-2}$ Hz$^{-1}$ sr$^{-1}$.

The full HST/FOS archival data set is presented in Paper III and 
can also be found at

\mbox{\tt http://lithops.as.arizona.edu/$\tilde{\;}$jill/QuasarSpectra} or

\mbox{\tt http://hea-www.harvard.edu/QEDT/QuasarSpectra},

\noindent so we describe the data used in this paper only briefly in \S\ref{sec-data}.
In \S\ref{sec-zsys} and \S\ref{sec-flux}, we discuss our treatment
of two parameters of each sample object which are integral to the proximity
effect analysis, systemic redshifts and Lyman limit fluxes.  We outline
the proximity effect analysis in \S\ref{sec-analysis} and we
present our results in \S\ref{sec-disc}.  We discuss the recovery
of $J(\nu_{0})$ from simulated Ly-$\alpha$ forest spectra in
\S\ref{sec-sims}.  A comparison of the results from radio loud and
radio quiet QSOs is given in \S\ref{sec-rl}.  The maximum likelihood
solutions for $J(\nu_{0})$ found when allowing for an equivalent
width threshold that varies across each sample spectrum 
are discussed in \S\ref{sec-varthr}. Solutions for
the HI ionization rate are given in \S\ref{sec-gam}.
The effect of a non-zero $\Omega_{\Lambda}$ on our
calculations is discussed in \S\ref{sec-omegal}, and 
the effect of the ionizing background on the Ly-$\alpha$ forest
line density is discussed in \S\ref{sec-dndz}.
We provide comparisons
with previous observational work on the low redshift UV background
in \S\ref{sec-prevres} and with models of this background in 
\S\ref{sec-models}.  A discussion of possible systematic effects on 
this analysis is given in \S\ref{sec-systematics}.  We conclude with
a summary of the results in \S\ref{sec-summary}.

\section{Data Sample}
\label{sec-data}

The reduction of the FOS data is described in Paper III.
Table~\ref{tab-z} lists the objects used in the proximity effect analysis
along with the object's redshift and classification in the NASA Extragalactic
Database.

For the reasons outlined in Scott et al.\ (2000b) we have removed from the
full FOS sample of Paper III
the spectra of quasars known to
be lensed, as well as those that show damped Ly-$\alpha$ absorption,
associated absorption, or broad intrinsic absorption.  
For our primary
proximity effect sample, we also remove objects classified as blazars
(BL Lacs and optically violent variables) on the grounds that their continua are highly 
variable.  However, we also perform the proximity effect analysis
with associated absorbers, damped Ly-$\alpha$ absorbers, and blazars included in order 
to determine if they affect the
results obtained.

\begin{figure}
\plotone{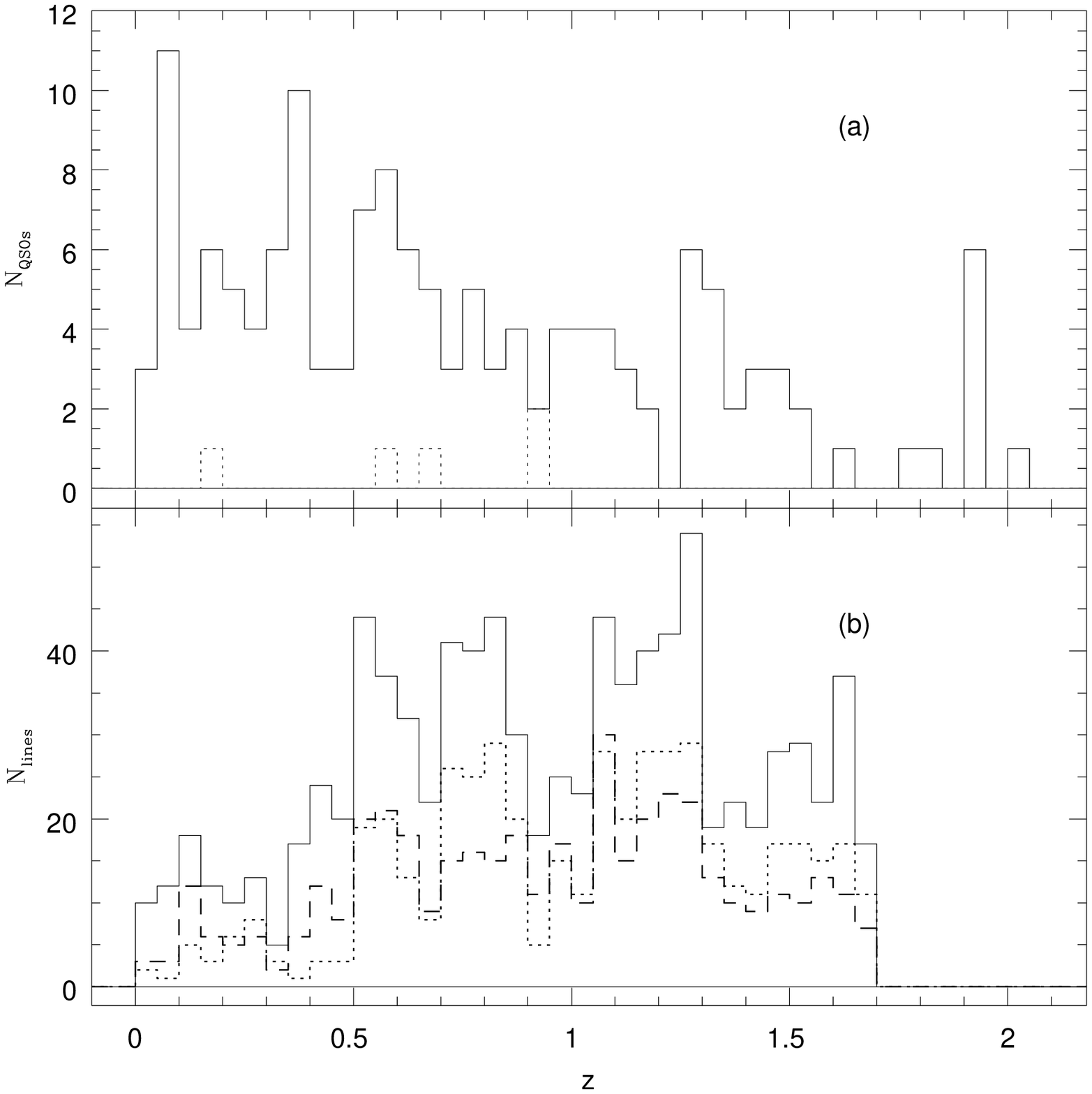}
\caption
{Histograms of (a) QSO redshifts in proximity effect sample, 
dotted line indicates objects classified as blazars or BL Lacs, and
(b) Ly-$\alpha$ line redshifts in proximity effect sample,  
(solid line)- lines above variable threshold,
(dashed line)- lines with $W > 0.32$~\AA\,
(dotted line)- lines with $W > 0.24$~\AA\ \label{fig-samphist}}
\end{figure}

As discussed in Paper III, objects observed only in the period before
the COSTAR upgrade to the HST optics and with the A-1 FOS aperture are
particularly subject to irregular line spread functions.  We have omitted those data
from this analysis as well.  The distributions in redshift of the QSOs and absorption lines
used in this paper are shown in Figure~\ref{fig-samphist}.

\section{Systemic Redshifts}
\label{sec-zsys}

QSO redshifts based on the Ly-$\alpha$ emission line have been shown to 
be blueshifted from the systemic redshift based on narrow emission lines
by up to $\sim 200$ km~s$^{-1}$.  Generally, the forbidden OIII doublet at 4959, 5007~\AA\ 
is taken to be the most reliable indicator of the QSO systemic redshift;  though
other lines such as Mg II $\lambda\lambda$2796,2803 and Balmer lines have been
shown to trace the systemic redshift as well, with some spread. 
(Zheng \& Sulentic 1990, Tytler \& Fan 1992, Laor et al.\ 1994,1995, Corbin \& Boroson
1996)
However,
the results of Nishihara et al.\ (1997), M$^{\rm c}$Intosh et al.\ (1999), and 
Scott et al.\ (2000b) indicate that in fact H$\beta$ may
not reflect the systemic redshift of high redshift QSOs. 

\subsection{Observations}
\label{sec-zobs} 
Spectra of the emission lines H$\beta$, [OIII]$\lambda$5007, or
Mg II were obtained for several objects in our total proximity effect sample.
The observations were carried out on the nights of
19 December 1995, 14 January 1996, 20 and 21 April 1996, 12 and 13 December
1996, and 2 February 1997.  These observations are summarized in Table~\ref{tab-zhst}.

The 19 December 1995 and 13 December 1996 observations were made using the
1.5 meter Tillinghast telescope at the Fred Lawrence Whipple Observatory
using the FAST spectrograph (Fabricant et al.\ 1998)
and a thinned Loral 512x2688 CCD chip (gain = 1.06,
read noise = 7.9 e$^{-}$)
binned by a factor of 4 in the
cross-dispersion direction.
Observations were made using a  300 lines mm$^{-1}$ grating blazed at
4750~\AA\ and a 3$\arcsec$ slit.
These spectra cover a wavelength range of
3660--7540~\AA.  This is listed as set-up (1) in Table~\ref{tab-zhst}.

The January, April, and December 10 and 12, 1996 observations
were made using the Steward Observatory
Bok 90 inch telescope using the Boller and Chivens
Spectrograph with a 600~l~mm$^{-1}$ grating blazed at 6681~\AA\ in the
first order, a 1.5$\arcsec$ slit, and
a 1200 x 800 CCD array with a gain of 2.2 e$^{-}$ ADU$^{-1}$ and a read noise of
7.7 e$^{-}$, binned 1x1.
For the January 1996 observations, the data were obtained with one of two grating tilts,
one resulting in wavelength coverages of 3600--5825~\AA\ and 
6870--9140~\AA.  
For the April 1996 data, the wavelength ranges were 
4140--6370~\AA\ and 5280--7550~\AA.
Two grating tilts were also used for the December 1996 data,
giving wavelength coverages of 4500--6700~\AA\ and  5610--7860~\AA.

The spectrum of one object, 0827+2421, was obtained
on 15 February 1997 at the Multiple Mirror Telescope with the
Blue Channel Spectrograph, a 2 $\arcsec$ slit,
the 3K x 1K CCD array, and the 800~l~mm$^{-1}$
grating blazed at 4050~\AA\ with spectral coverage of 4365--6665~\AA.

\begin{figure}
\plotone{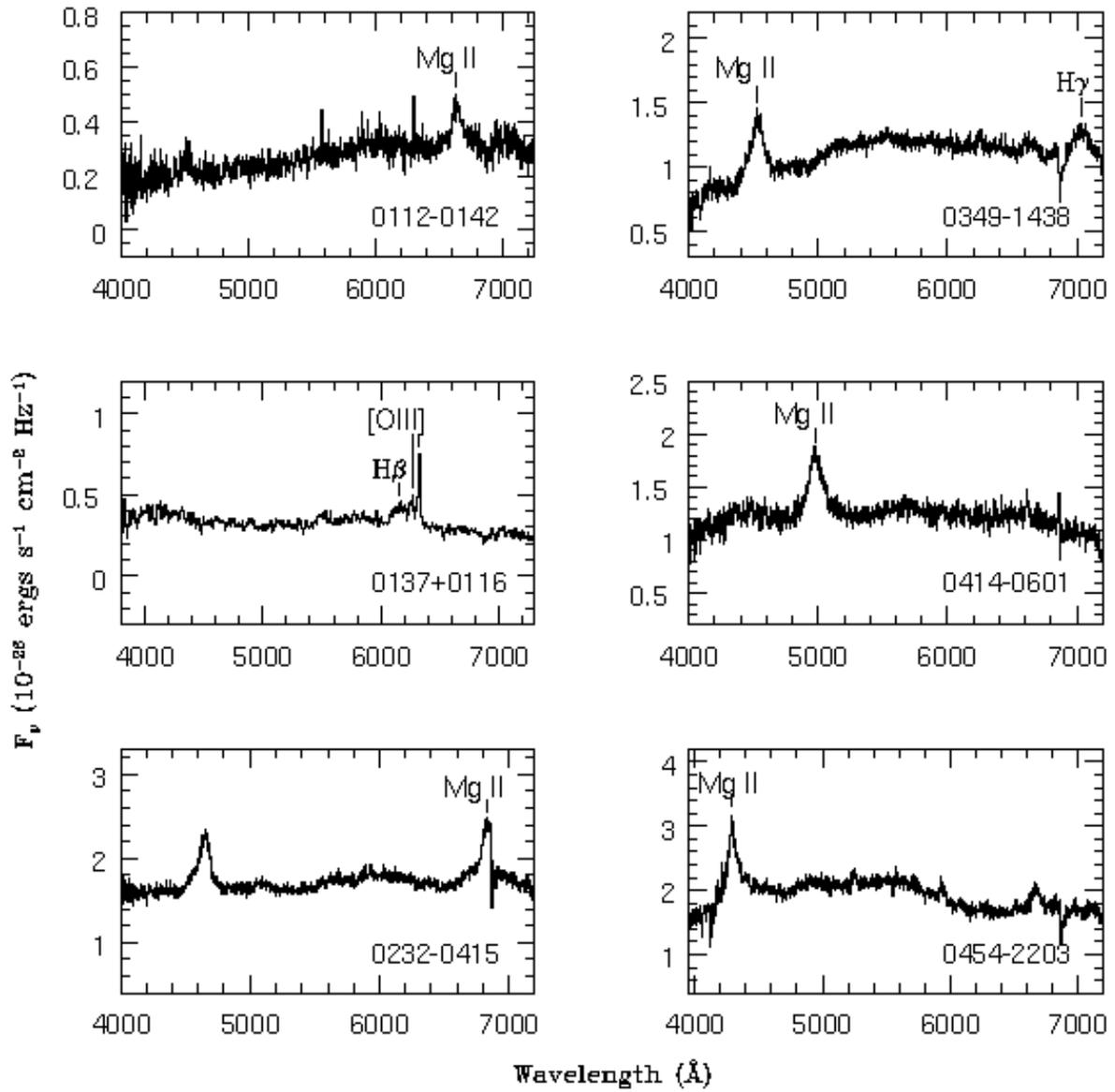}
\caption
{Emission line spectra of sample QSOs used to measure redshifts
\label{fig-zspec}}
\end{figure}

The spectra are shown in Figure~\ref{fig-zspec} and the lines used for
redshift measurements are labeled.

\subsection{Measurements}
\label{sec-zmeas}
Taking a simple cursor measurement of each line centroid, we find
a mean  [OIII]-Balmer line $\Delta$v of -30 $\pm$ 1010 km~s$^{-1}$ 
for 31 objects and a mean [OIII]-Mg II $\Delta$v of 58 $\pm$ 576 km~s$^{-1}$
for 31 objects.  The mean blueshift of the Ly-$\alpha$ emission line with
respect to [OIII] is 289 $\pm$ 727 km~s$^{-1}$ based on 51 measurements.
The redshifts measured for each object in our sample are shown in Table~\ref{tab-z}; and
the results are shown in Figure~\ref{fig-zhist}.
Gaussian fits to the lines give similar results.

We therefore treat both Balmer lines and Mg II in addition to [OIII] as good systemic redshift
indicators for these low redshift objects.   In the case of a QSO for which we have
only a Ly-$\alpha$ emission line measurement of the redshift, we add 
300 km~s$^{-1}$ to this value to estimate its systemic redshift. 

\section{Lyman Limit Fluxes}
\label{sec-flux}

Our method for estimating Lyman limit fluxes for each QSO is the same as 
that described in Paper II.  For objects with spectral coverage between
the Ly-$\alpha$ and CIV emission lines,
we extrapolate the flux from 1450~\AA\ in the
quasar's rest frame to 912~\AA\ using $f_{\nu} \sim \nu^{-\alpha}$ and
a spectral index $\alpha$ measured primarily from the
spectral region between the Ly-$\alpha$ and C IV emission lines.
Figure~\ref{fig-fnus}
shows the FOS spectra for which these fits were made along with the
power law fits themselves.  In some cases, $\alpha$ is poorly
constrained from these fits, especially if there was
little spectral coverage redward of Ly-$\alpha$ emission in the data. 
If another measurement of the spectral index was available in the literature for 
these objects, we used it; otherwise, we used our measurement.

Table~\ref{tab-flux} lists the Lyman limit flux for each object in this
proximity effect sample and either a) the flux at 1450~\AA, or some
other appropriate wavelength free of emission features, measured
from the FOS data, or b) a directly
measured Lyman limit flux and the reference.
If available from the extracted archive data, red spectra
and the fits to them are presented for objects which were observed only with pre-COSTAR FOS
and A-1 aperture, though these data were not subsequently used for any Ly-$\alpha$
forest studies. See Table 4 of Paper III. 

In Figure~\ref{fig-zL}, we show QSO Lyman limit luminosities versus emission
redshift for this HST/FOS sample
combined with the high redshift objects presented in Papers I and II.  Only
at the lowest redshifts is there any trend of luminosity with redshift. 

\section{Analysis}
\label{sec-analysis}

The distribution of Ly-$\alpha$ lines in redshift and equivalent width is
given by:
\begin{equation}
\frac{\partial^{2} {\cal N}}{\partial z \partial W}= \frac{A_{0}}{W^{*}}(1+z)
^{\gamma}\exp\left(-\frac{W}{W_*}\right)  \label{eq:dndzdw}.
\end{equation}
The distribution in redshift and HI column density, N, is:
\begin{equation}
\frac{\partial^{2} {\cal N}}{\partial z \partial N}= A N^{-\beta}
(1+z)^{\gamma}
\label{eq:dndzdnh1}.
\end{equation}
The parameter
$\gamma$ is the redshift distribution parameter. 
The quantities $W^{*}$ in Equ.~\ref{eq:dndzdw} and
$\beta$ in Equ.~\ref{eq:dndzdnh1} are the line rest equivalenth width
and column density distribution parameters, respectively.
The quantities $A_{0}$ and $A$ are normalizations.

The BDO method for measuring $J(\nu_{0})$ consists of 
binning all lines in the sample in the parameter $\omega(z)$,
the ratio of
QSO to background Lyman limit flux density at the physical location
of the absorber:
$F^{Q}(\nu_{0})/ (4 \pi J(\nu_{0})) \label{eq:omega}$ 
for various values of $J(\nu_{0})$.  The value of 
$J(\nu_{0})$ that results in the lowest $\chi^{2}$ between the
binned data  and the ionization model, 
\begin{equation}
\frac{d{\cal N}}{dz}= {\cal A}_{0}(1+z)^{\gamma}[1+\omega(z)]^{-(\beta-1)},
\label{eq:dndx}
\end{equation}
is considered to be the optimal value.
This ionization model follows from the assumption that
the column densities of lines are modified by the presence
of the QSO according to 
\begin{equation}
N \propto N_{0}(1+\omega(z))^{-1},
\label{eq:column}
\end{equation}
where $N_{0}$ is the column density a given line would
have in the absence of the QSO.
The 1$\sigma$ errors are found from 
$\Delta \chi^{2}=8.18$ for 7 degrees of freedom (Press et al.\ 1992).

The value of $\omega(z)$ for each line in a given sample
depends not only upon the value of $J(\nu_{0})$ assumed, but
also on the cosmological model, as
\begin{equation}
F^{Q}(\nu_{0})= \frac{L(\nu_{0})}{4\pi r_{L}^{2}(z)}
\label{eq:fq}
\end{equation}
and
\begin{equation}
L(\nu_{0}) = 4\pi d_{L}^{2}(z) \frac{f(\nu_{0})}{(1+z_{em})},
\end{equation} 
where $r_{L}(z)$ is the luminosity distance of an individual absorber from the
QSO and $d_{L}(z)$ luminosity distance to the QSO from the observer.  
The luminosity distance between two objects at different redshifts
can be calculated analytically  for
cosmological models in which $\Omega_{\Lambda}=0$.  We return
to this point in Section~\ref{sec-omegal} below.

If the proximity effect is indeed caused by enhanced ionization of the IGM in the
vicinity of QSOs, one may expect to observe a larger deficit of lines relative
to the Ly-$\alpha$ forest near high
luminosity QSOs than near low luminosity QSOs.
In Figure~\ref{fig-npred}(a), we plot the fractional deficit of lines
with respect to the number predicted by Equ.~\ref{eq:dndzdw} versus
distance from the QSO for this HST/FOS sample combined with the
high redshift objects observed with the Multiple Mirror Telescope (MMT) 
presented in Papers I and II. 
We divide our QSO sample into high and low luminosity objects at 
the median Lyman limit luminosity of the combined MMT and HST/FOS sample, 
log(L$_{{\rm 912} \; \mbox{\scriptsize\AA}}$)~$\sim$~31.
High luminosity objects show a marginally more
pronounced proximity effect than low luminosity objects:  4.9$\sigma$
for QSOs with log(L$_{{\rm 912} \; \mbox{\scriptsize\AA}}$)~$>$~31 versus 3.2$\sigma$ for QSOs
with log(L$_{{\rm 912} \; \mbox{\scriptsize\AA}}$)~$<$~31. In panel (b), we plot the line 
deficit within 2 h$_{75}^{-1}$ Mpc as a function of log(L$_{{\rm 912} \; \mbox{\scriptsize\AA}}$).
The lack of a significant difference in the line deficit between high and low luminosity QSOs
may indicate the presence of clustering, if absorption features cluster more strongly around
more luminous QSOs with deeper potential wells.
We will address the issue of clustering further below.

The BDO method of measuring the background can result in poor statistics
at low redshift due to the low line density in the low redshift Ly-$\alpha$ forest.
We will quote results from this method, but we will generally
the maximum likelihood method for measuring $J(\nu_{0})$ as presented by
KF93, which consists of constructing a 
likelihood function of the form
\begin{equation}
L= \prod_{a}{\it f}(N_{a},z_{a}) \prod_{q} {\rm exp}[ - \int^{z^{q}_{\rm max}}
_{z^{q}_{\rm min}} dz \int^{\infty}_{N^{q}_{\rm min}} {\it f}(N,z) dN ] \label{eq:maxlike},
\end{equation}
where
\begin{equation}
{\it f}(N,z) = AN^{-\beta}(1+z)^{\gamma}[1+\omega(z)]^{-(\beta-1)},
\label{eq:fnz}
\end{equation}
and the indicies $a$ and $q$ denote sample absorption lines and quasars, respectively.
Using the values of $\gamma$ and $A_{0}$ from a separate maximum likelihood
analysis on the Ly-$\alpha$ forest excluding regions of the spectra 
affected by the proximity effect (Dobrzycki et al. 2001, Paper IV), 
and a value of $\beta$ from studies with high resolution data, eg.
$\beta=1.46$ from Hu et al.\ (1995), the search for the best-fit value of
$J(\nu_{0})$ consists of finding the value that maximizes this function,
fixing the other parameters. 

If the line density is low throughout a single  Ly-$\alpha$ forest spectrum,
it becomes difficult to distinguish any proximity effect, even
in a large sample of spectra.  The absence of a proximity effect in this
model formally translates into the limit $J(\nu_{0}) \rightarrow \infty$ 
because in this scenario,
the QSO has no additional effect on its surroundings and therefore generates
no relative line underdensity.  The errors quoted in the values of log[$J(\nu_{0})$]
are found from the fact that in solving for log[$J(\nu_{0})$] alone,
the logarithm of the likelihood function,
$-2 {\rm ln}(L/L_{\rm max})$, is distributed as $\chi^{2}$ with one degree
of freedom.  In the case of an ill-defined solution, the likelihood
function is very broad and the formal error approaches infinity.  
If a proximity effect is weak but not absent in the data, 
a maximum likelihood solution is sometimes possible, but with no well-defined
1$\sigma$ upper limit on log[$J(\nu_{0})$].  In other words, if an upper limit of 
infinity is quoted, the data cannot
rule out the nonexistence of a proximity effect to within 1$\sigma$ confidence.

Using a constant equivalent width threshold results in the loss of 
a large amount of spectral information. 
In the case of a large equivalent width threshold, of course,
many weak lines are discarded; 
and in the case of a small threshold, regions of spectra 
where the signal-to-noise ratio (S/N) does not permit the detection of lines all
the way down to the specified threshold are lost and 
only the highest S/N spectral regions are used. 
The technique of measuring the statistics $\gamma$ and $W^{*}$
has been expanded to allow for a threshold that varies with
S/N  across each QSO spectrum
(Bahcall et al.\ 1993,1996, Weymann et al.\ 1998,
Scott et al.\ 2000a).   We will use this variable threshold information
to measure $J(\nu_{0})$ as well.

\section{Results}
\label{sec-results}

The results of this analysis are given in Table~\ref{table-jnu}.

Before we begin the discussion of the results, some words about
the normalization values listed in Table~\ref{table-jnu} are in order.
In the BDO method for measuring $J(\nu_{0})$, lines are binned in $\omega(z)$ and
compared to the ionization model given by  Equ.~\ref{eq:dndx}, for
an assumed value of $\beta$.  
In this case, the normalization listed in Table~\ref{table-jnu} is
the parameter in Equ.~\ref{eq:dndzdw}, found from the number of
lines in the sample and the 
maximum likelihood value of $\gamma$:
\begin{equation}
{\cal A}_{0} =  A_{0}\exp\left(-\frac{W_{\rm lim}}{W_*}\right) =
{\cal N} \left( \sum_{q} \int_{z_{\rm min}^{q}}^{z_{\rm max}^{q}} dz \, (1+z)^{\gamma} \right)^{-1},
\label{eq:bdonorm}
\end{equation}
where ${\cal N}$ is the total number of lines observed with rest equivalent width greater than
$W_{\rm lim}$, the limiting equivalent width of the line sample.
For the maximum likelihood solutions for $J(\nu_{0})$, we convert line
equivalent widths to column densities using the Ly-$\alpha$ curve of growth 
and an assumed value of $b$, the characteristic Doppler parameter of the lines.
As we will demonstrate, different values of $\beta$ and $b$
have only a small effect on the value of $J(\nu_{0})$ found.
The normalization is given by
\begin{equation}
A =
{\cal N} \left( \sum_{q} \int_{z_{\rm min}^{q}}^{z_{\rm max}^{q}} dz 
\int_{N_{\rm min}^{q}(z)}^{\infty} dN \, N^{-\beta} (1+z)^{\gamma} \right)^{-1},
\label{eq:mlnorm}
\end{equation}
where $N_{\rm min}^{q}(z)$  is the limiting column density across each
QSO spectrum corresponding to a limiting equivalent width.  
This quantity can be held constant, as in the BDO method, or it can be
allowed to vary across each QSO spectrum.
In both of these formulations for the normalization, a proximity
region around the QSO is neglected and that proximity region is
either defined by a velocity cut, eg. $z_{\rm em}$ - 3000 km~s$^{-1}$,
or by a cut in $\omega(z)$, eg. $\omega(z)=0.2$.  

We also use the standard BDO method to find
$\log[J(\nu_{0})]=
-22.04^{+0.43}_{-1.11}$ and
$-22.06^{+0.05}_{-0.62}$
for equivalent width thresholds of 0.32 and 0.24~\AA\
respectively.
Figures~\ref{fig:chi2}(a) and (d) illustrate the $\chi^{2}$ of the
binned data compared to the BDO ionization model
as a function of assumed $J(\nu_{0}$) for these two thresholds.
The BDO ionization model is expressed in terms
of the number of lines per coevolving coordinate:
\begin{equation}
\frac{dN}{dX_\gamma} = {\cal A}_{0} (1 + \omega)^{-(\beta - 1)},
\label{equ:dndx2}
\end{equation}
where $X_\gamma= \int (1+z)^{\gamma} dz$.
This $\chi^{2}$ curve
is very broad, which is reflected in the large
error bars and indicates the difficulty in
isolating the optimal mean intensity of a weak
background using this technique. Figures~\ref{fig:dndx}(a) and (d)
show the binned data and the ionization model for the
values of $J(\nu_{0}$) listed above, those
that give the lowest $\chi^{2}$ between the binned
data and the model, ie. the minima of the curves in
Figures~\ref{fig:chi2}(a) and (d).

We executed the maximum likelihood search for $J(\nu_{0})$,
using two different fixed equivalent width thresholds,
0.24~\AA\ and  0.32~\AA\, as well as for the case of a 
variable threshold across all the spectra.  
The uncertainty in $\gamma$ does not translate directly into a large uncertainty
in $J(\nu_{0})$.
Changing the value of $\gamma$ alters the maximum likelihood normalization, $A$, according
to Equ.~\ref{eq:mlnorm}.  From the sample of lines with rest equivalent widths greater
than 0.32 \AA\, we find $\log[J(\nu_{0})]= -22.11^{+0.51}_{-0.40}$ for
$\gamma=0.82 \pm 0.29$.  Varying $\gamma$ by $\pm 1\sigma$ gives
$\log[J(\nu_{0})]= -22.21$
and $-22.00$ with similar uncertainties.

The data used here are not of sufficient resolution to fit Voigt profiles
to the absorption features and derive HI column densities and
Dopper parameters.  We therefore choose
to fix the values of $\beta$ and $b$ to those found from work on high resolution data,
rather than
allow them to freely vary in our analysis.
For the 0.32~\AA\ 
fixed equivalent width threshold,
we tested several pairs of values of $(\beta,b)$ where
$b$ is in km~s$^{-1}$:
(1.46,35) and  (1.46,25) where the value of
$\beta$ is taken from Hu et al.\ (1995); as well as (1.45,25) and (1.70,30)
found from low redshift Ly-$\alpha$ forest spectra taken
with the Goddard High Resolution Spectrograph (GHRS) on HST by 
Penton et al.\ (2000a,b).  In addition, Dav\'{e} \& Tripp (2001) have found some evidence
for $\beta$ increasing to 2.04 at $z < 0.3$ from high resolution echelle
data from the Space Telescope Imaging Spectrograph aboard the HST.  We
test this value as well.
The likelihood functions for the
maximum likelihood solutions listed in rows 2-6, 8-12, 14, and 18 of
Table~\ref{table-jnu} are shown in 
Figure~\ref{fig:like1}.  The binned data and ionization
models are plotted in Figure~\ref{fig:dndxl1}.
The values of $J(\nu_{0})$ derived for these
various pairs of values of $\beta$ and $b$ are not significantly
different from one another, though the results in 
Table~\ref{table-jnu} indicate that varying $\beta$ has a larger
impact on the inferred $J(\nu_{0})$ than does varying $b$.  The solution
for $\beta=2.04$ differs from the $\beta=1.46$ solution by $\sim 1\sigma$.
In the analysis that follows,
we adopt the values 1.46 and 35 km~s$^{-1}$. 

The models
of Haardt \& Madau (1996) predict that the UV background arising
from QSOs drops by over an order of magnitude from $z=2.5$ to $z=0$. 
We therefore 
divide the sample into low and high redshift subsamples at $z=1$ and
use both the BDO method and the maximum likelihood method for finding
$J(\nu_{0})$.  
These results, also listed in Table~\ref{table-jnu},
confirm some evolution in $J(\nu_{0})$, though not at a high level of significance. 
For the BDO solutions,
we find log[$J(\nu_{0})$] at $z < 1$ is equal to $-22.87^{+1.19}_{-0.82}$
and log[$J(\nu_{0})$] at $z > 1$ is equal to $-22.02^{+0.005}_{-1.33}$.
The restrictive 1$\sigma$ upper limit for log[$J(\nu_{0})$] at $z > 1$
arises from the steeply rising
$\chi^{2}$ as a function of log[$J(\nu_{0})$] shown in Figure~\ref{fig:chi2}. 
This, in turn arises from the single line in the highest $\log(\omega)$ bin
moving to the next bin for larger values of $J(\nu_{0})$, resulting in 
a drastic change in the $\chi^{2}$ with respect to the photoionization model.  
We do not consider this to be a reliable indicator of the uncertainty in $J(\nu_{0})$ at $z > 1$. 
The maximum likelihood technique gives more robust estimates of the uncertainties.
From this analysis, we find
log[$J(\nu_{0})$] at $z < 1$ is found to be $-22.18^{+0.90}_{-0.61}$, while
at $z > 1$ it is -21.98$^{+0.76}_{-0.54}$.
These results are shown in Figures~\ref{fig:lowzcomp}(a) and 
\ref{fig:allzcomp}.

Including associated absorbers, damped Ly-$\alpha$ absorbers, or blazars
in the proximity effect analysis appears to have little effect on
the results.  One might expect associated absorbers to
reduce the magnitude of the observed proximity effect and hence
cause $J(\nu_{0})$ to be overestimated.  The value found including
the 45 associated absorbers in our sample is indeed larger, 
log[$J(\nu_{0})$]$=-21.74^{+0.55}_{-0.39}$, versus
log[$J(\nu_{0})$]$=-22.11^{+0.51}_{-0.40}$, but not significantly so.
Likewise, if the intervening dust extinction in damped Ly-$\alpha$ absorbers 
is significant, including these objects in our analysis could 
cause us to overestimate the magnitude of the proximity effect and
hence underestimate $J(\nu_{0})$.  However, the inclusion of these 7
objects only negligibly reduces the value of $J(\nu_{0})$ derived.
QSO variability on timescales less than $\sim 10^{5}$ years would be expected 
to smooth out the proximity effect distribution (BDO).  However, the inclusion
of 6 blazars in the sample, all at $z < 1$, resulted in no discernible change in $J(\nu_{0})$.
The sample used in the analysis of HI ionization rates discussed below
includes all of these objects. 

For each solution, we calculate the $\chi^{2}$ with respect to the
ionization model expressed by Equ.~\ref{eq:dndx},
and the probability that the observed $\chi^{2}$ will
exceed the value listed by chance for a correct model, Q$_{\chi^{2}}$ 
(Press et al.\ 1992).
We also execute a Kolmogorov-Smirnov (KS) test for each solution.
The KS test provides a measure of how well the assumed parent
distribution of lines with respect to redshift,
given by Equ.~\ref{eq:fnz}, reflects the true redshift distribution of lines
(cf. Murdoch et al.\ 1986, Press et al.\ 1992).
The KS probability, Q$_{\rm KS}$, indicates the probability that
a value of the KS statistic larger than the one calculated could have
occurred by chance if the assumed parent is correct. 
The KS probability associated with each solution for $J(\nu_{0})$
is listed in column 10 of Table~\ref{table-jnu}. 

\subsection{Simulations}
\label{sec-sims}

We tested our maximum likelihood methods, including
our treatment of the variable equivalent width thresholds by running
our analysis on a simulated data set.  Each of the 151 
spectra in this simulated data set had a redshift
equal to that of an object in our data set. All objects
including those showing associated absorption, damped Ly-$\alpha$ absorption,
or blazar activity  are included in this simulated set.
Each spectrum is created using a Monte Carlo technique
by which lines are placed in redshift and column density space
according to Equ.~\ref{eq:dndzdnh1}.
A background of known mean intensity modifies the
column densities of the lines 
according to the BDO formulation given
by Equ.~\ref{eq:column}.
The same analysis done on the data, consisting of the
line-finding algorithm and the maximum likelihood searches
for $\gamma$ and $J(\nu_{0})$,
is then used
on the simulated spectra in order to recover the
input $J(\nu_{0}$).  Three different values of log[$J(\nu_{0})$] are input,
-21, -22, and -23, and the results are listed in Table~\ref{table-sim}.
In order to understand the possible range of recovered log[$J(\nu_{0})$], 
we repeated the input log[$J(\nu_{0})$]$=-22$ simulation in the 
constant threshold case nine additional times,
resulting in $\overline{J(\nu_{0})}= 2.91 \pm 1.67 \times 10^{-22}$
ergs s$^{-1}$ cm$^{-2}$ Hz$^{-1}$ sr$^{-1}$.
In addition, since we observe the background to
evolve with redshift from $z=1$ to $z=0$, we implement a model
in which $J(\nu_{0})$ varies as a power law in $(1+z)$ over the redshift
range of the data.  This relationship is defined by the best fit to 
a power law variation of $J(\nu_{0})$ with redshift: 
$\log[J(\nu_{0})]= 0.017\log(1+z)-21.87$.
We recover this using both
the constant threshold and the variable threshold analyses, at all redshifts and
at $z < 1$ and $z > 1$ separately.
The results of this exercise
are shown in Table~\ref{table-sim} and in Figure~\ref{fig:sims}.

These simulation results indicate that both the constant and variable threshold
analyses can overestimate the background by up to a factor of 3-5, though
the uncertainties for the variable threshold solutions are consistently
lower, as a factor of $\sim 2$ more lines are used in these solutions.
We separated the first of the input log[$J(\nu_{0})$]$=-22$
simulated data samples into high and low redshift subsamples
at $z=1$,
in order to determine if the change in $J(\nu_{0})$ as 
a function of redshift could be falsely introduced in a case
there the input background is constant with redshift.  For 
both the constant and variable threshold treatments, this
is not the case.  The value found for the
low redshift subsample is actually larger than the value 
found for the high redshift subsample in both treatments.

In the case of the varying input $\log[J(\nu_{0})]$, 
the values recovered for the high redshift subsample
and for the entire redshift range of the data are overestimates.
The slope of the linear relationship between  $\log[J(\nu_{0})]$ and
$\log(1+z)$ is quite small, 0.017, resulting in a
variable input $\log[J(\nu_{0})]$ that is actually nearly constant with
redshift. 
The solution for $z < 1$
matches the input well for both the constant and variable threshold cases.
At $z > 1$, the variable threshold solution overestimates the input
by a larger factor, $\sim$3, or 1.6$\sigma$, than does the constant
threshold solution, $\sim$2, or less than 1$\sigma$.

In Paper II, we argued
that curve-of-growth effects are likely to come into play in
the proximity effect analysis and to play a larger role
for cases in which $J(\nu_{0})$ is large and the proximity
effect signature is small.  
Here we find that the input $J(\nu_{0})$ is recovered most 
effectively by the constant
and variable threshold cases for the largest input value of $\log[J(\nu_{0})]$, $-21$.
However, nearly every case tested with these simulations results in a value
of $J(\nu_{0})$ larger than the input value, especially when a variable equivalent
width threshold is used.
The only case where the difference is
significant is the input $\log[J(\nu_{0})]=-23$, variable threshold case.  The recovered
$\log[J(\nu_{0})]$, -22.47, is  4$\sigma$ larger than the input.   We will return to the
discussion of the variable threshold in Section~\ref{sec-varthr} below.

\subsection{HI Ionization Rate}
\label{sec-gam}

As described in Paper II, solving for the HI ionization rate,
\begin{equation}
\Gamma=\int_{\nu_{0}}^{\infty} \frac{4 \pi J(\nu) \sigma_{HI}(\nu)}{h \nu} d\nu
\; \; {\rm s^{-1}},
\label{eq:gamma}
\end{equation}
instead of $J(\nu_{0})$ avoids the
assumption that the spectral indicies of the QSOs and
the background are identical.  We modified our maximum likelihood
code to use the values of $\alpha$ for each QSO listed in 
Table~\ref{tab-flux} to measure this quantity and the
results are listed in Table~\ref{table-gamma}.  
For objects with no available measured value of $\alpha$, we use
$\alpha=2.02$, the extreme ultraviolet spectral index 
measured from a composite spectrum of 101 HST/FOS QSO spectra by
Zheng et al.\ (1997).  The result for lines above a constant 0.32 \AA\
rest equivalent width threshold is $\log(\Gamma)=-12.17^{+0.50}_{-0.40}$.
This result is not substantially changed 
if we instead use $\alpha=1.76$, the value found from a composite of
184 QSO spectra from HST/FOS, GHRS, and STIS by Telfer et al.\ (2001), giving
$\log(\Gamma)=-12.25^{+0.47}_{-0.35}$.
We also find little change in the result if we assume $\alpha=2.02$ or $\alpha=1.76$ for all QSOs.
The variable threshold data result
in a high HI ionization rate, and this is discussed further in the following section.
The constant threshold result is plotted in
Figure~\ref{fig:gam}.
Evolution in the UV background is more apparent in the HI
ionization rate than in the solutions for $J(\nu_{0})$.
The result at $z > 1$ is 6.5 times larger than  that at $z < 1$.
The values of $J(\nu_{0})$ implied by these solutions for $\Gamma$ and  a global source
spectral index
$\alpha_{\rm s}=1.8$ are also listed
in Table~\ref{table-gamma}.

We also parametrize the evolution of the HI ionization rate
as a power law:
\begin{equation}
\Gamma(z)= A_{\rm pl}(1+z)^{B_{\rm pl}}
\label{equ:plgam}
\end{equation}
and solve for the parameters $A_{\rm pl}$ and $B_{\rm pl}$ in both the constant and variable threshold
cases.  The values we find are shown as the dashed line in Figure~\ref{fig:gam}
also listed in Table~\ref{table-gamma}. 

HM96 parametrize their models of the HI ionization rate with the function:
\begin{equation}
\Gamma(z)= A_{\rm HM}(1+z)^{B_{\rm HM}} \exp \left( \frac{-(z-z_{c})^{2}}{S} \right)
\label{equ:hmgam} 
\end{equation}
We combine our data set with that of Scott et al.\ 2000b to solve for the
parameters $A_{\rm HM}$, $B_{\rm HM}$, $z_{c}$, and $S$.  We find $(A_{\rm HM},B_{\rm HM},z_{c},S)$=
($7.6 \times 10^{-13}$, 0.35, 2.07, 1.77)
for $\beta$=1.46 and 
$(A_{\rm HM},B_{\rm HM},z_{c},S)$~=~($3.2 \times 10^{-13}$, 1.45, 2.13, 1.42) for $\beta$=1.7,
while the
parameters found by HM96 for $q_{0}=0.5$ are ($6.7 \times 10^{-13}$, 0.43, 2.30, 1.95).
These results are also represented by the solid curves in Figure~\ref{fig:gam}, while
the HM96 parametrization is shown by the dotted line for comparison.   

\subsection{Variable Equivalent Width Threshold}
\label{sec-varthr}

The variable threshold analysis yielded some unexpected results.
As seen in the majority of the simulations,
the values of $J(\nu_{0}$) found were consistently larger than the values
found using a constant equivalent width threshold, indicating
that the inclusion of weaker lines suppresses the proximity
effect.  This is to be expected if clustering is occurring (Loeb \& Eisenstein 1995),
which in itself is to be expected to be more prominent at low
redshift than at high redshift.  However, the suppression of
the proximity effect by the inclusion of weak lines is somewhat counterintuitive
from the perspective of the curve of growth.  Most of the lines included in
a constant threshold solution are on the flat part of the curve
of growth.  Therefore, though the ionizing influence of the quasar
may be translated directly into a change in the HI column density,
as predicted by the BDO photoionization model, this will not necessarily
result in a corresponding change in the line equivalent width.
The solution for $z < 1$ is nearly a factor of 3 larger than the
the solution found in the case of a constant, 0.32~\AA\ equivalent width threshold.
The solution for $z > 1$ is a factor of $\sim 6$ larger than the constant threshold solution,
with no well-defined 1$\sigma$ upper limit
due to the flattening of the likelihood function towards high $J(\nu_{0}$)
This likelihood function for the total sample
shows two peaks,
the most prominent at log[$J(\nu_{0}$)]$=-20.82$, the solution listed
in Table~\ref{table-jnu}, and a secondary peak at log[$J(\nu_{0}$)] $\sim -18.4$.

This behavior is also exhibited, even more dramatically,
in the solutions for the HI ionization rate, as
discussed above.
We conducted a jackknife resampling experiment (Babu \& Feigelson 1996, Efron 1982)
to determine the source of these likelihood function peaks at large log($\Gamma$), or
log[$J(\nu_{0}$)].

Two objects, 0743-6719 ($z_{\rm em}= 1.508$) and 0302-2223 ($z_{\rm em}= 1.402$),
are found from jackknife experiments to produce
all of this effect.  In the jackknife experiment, we perform the
maximum likelihood calculation of $J(\nu_{0}$) N times, where N
is the number of objects in the high redshift subsample.  In each calculation, one
object from the total sample is removed.  The results of this experiment
are shown in the histogram in
Figure~\ref{fig:jack}.
The removal of 0743-6719 or 0302-2223 results in
the two values of $\Gamma$ that are well-defined
and that are in
reasonable agreement with the value calculated at high redshift in the
constant threshold case.
Removing only the one line from 0743-6719
nearest the Ly-$\alpha$ emission line with $z_{\rm abs}=1.5058$ and
observed equivalent width equal to 0.23~\AA\, results in
$\Gamma = 6.23 \times 10^{-12}$ s$^{-1}$.
This object was part of the HST Key Project sample (Jannuzi et al.\ 1998)
and they cite no evidence of associated aborption in its spectrum.
Removing only the one line from 0302-2223
nearest the Ly-$\alpha$ emission line with $z_{\rm abs}= 1.3886$ and observed
equivalent width equal to 0.27~\AA\, results in
$\Gamma = 8.14 \times 10^{-12}$ s$^{-1}$.
This object shows an absorption system at $z_{\rm abs}=1.406$ and is
classified as an associated absorber.  No metal absorption is seen at
$z_{\rm abs}= 1.3886$, though this absorber is within 5000 km~s$^{-1}$
of the QSO, the canonical associated absorber region.
Removing both of these lines gives $\Gamma = 3.88 \times 10^{-12}$ s$^{-1}$.
Due to the small equivalent  widths of both of these
lines they are not included in the constant threshold analysis, and the
solutions for $J(\nu_{0})$ and $\Gamma$ for $z > 1$ are well-defined.

It appears that this method
has some trouble reliably recovering the background from a sample
of absorption lines above an equivalent width threshold allowed to vary
with S/N.
As the method works well for the constant threshold case, we contend that 
the photoionization model, expressed in  Equ.~\ref{eq:dndx}, used to create the likelihood function
must not be an adequate model for the proximity effect when weak lines
are included in the analysis.   Liske \& Williger (2001) introduce a method 
for extracting $J(\nu_{0})$ from QSO spectra based on flux statistics.
We shall return to this topic in future work.

\section{Discussion}
\label{sec-disc}

\subsection{Radio Loudness}
\label{sec-rl}

As the results listed in Table~\ref{table-jnu} indicate, the inclusion of the
four blazars and one BL Lac object, all at $z < 1$, in our sample
does not change the result significantly.  However,
there is much observational evidence that radio loud and radio quiet
quasars inhabit different environments, namely that radio loud quasars
reside in rich clusters while radio quiet quasars exist in galaxy environments
consistent with the field
(Stockton 1982, Yee \& Green 1984, 1987, Yee 1987, Yates, Miller, \& Peacock 1989,
Ellingson, Yee, \& Green 1991, Yee \& Ellingson 1993,
Wold et al.\ 2000, Smith, Boyle, \& Maddox 2000).
If there is a corresponding increase in
the number of Ly-$\alpha$ absorption lines in the spectra of radio loud objects,
this could cause the proximity
effect to be suppressed, and the measured log[$J(\nu_{0})$] to be artificially
large.
We have therefore divided our sample
into radio loud and radio quiet subsamples using the ratio of radio to
UV flux to characterize the radio loudness,
\begin{equation}
{\rm RL =  log[S(5 \; GHz)]/log[S(1450} \; \mbox{\AA})].
\end{equation}
The value of RL for each object in our sample is listed in Table~\ref{tab-flux}.
A histogram of these values and the distribution of RL with $z$ for the sample objects
are shown in Figure~\ref{fig:rl}. The division between radio loud and radio
quiet was chosen to be RL=1.0.  The resulting values of log[$J(\nu_{0})$] for
these subsamples are listed in Table~\ref{table-jnu}.  There is no significant
trend for log[$J(\nu_{0})$] to appear larger for radio loud objects than for
radio quiet objects.

\subsection{Non-Zero $\Omega_{\Lambda}$}
\label{sec-omegal}

We performed the maximum likelihood calculation for the
case of a non-zero cosmological constant.  This means that
the  observer-QSO and absorber-QSO luminosity distances
that appear in the
relationship between $\omega$ and $z$ (BDO)
must be calculated numerically from the expression:
\begin{equation}
d_{L} = (1+z) \frac{c}{H_{0}} \int_{0}^{z} \frac{dz^{\prime}}{E(z^{\prime})}, 
\end{equation}
where
\begin{equation}
E(z) \equiv \sqrt{\Omega_{\rm M}(1+z)^{3} + \Omega_{\rm k}(1+z)^{2} + \Omega_{\Lambda}},
\end{equation}
(Peebles, 1993) as this integral cannot be reduced to an analytical form for
$\Omega_{\Lambda} \neq 0$.

The calculations in the sections above assume 
($\Omega_{\rm M}$,$\Omega_{\Lambda}$)~=~(1.0,0.0).
Here, we perform the maximum likelihood search for $J(\nu_{0}$) using 
($\Omega_{\rm M}$,$\Omega_{\Lambda}$)~=~(0.3,0.7).
For a QSO at $z=0.5$ with a Lyman limit flux density of
0.1 $\mu$Jy, an absorber at $z=0.48$, and an assumed background of log[$J(\nu_{0})$]$=-22$., 
this ($\Omega_{\rm M}$,$\Omega_{\Lambda}$) results in
a value of $\omega$ that is $\sim 25$\% smaller than that inferred in the 
$\Omega_{\Lambda}=0$ case. 
Unlike all the other solutions performed, we ignore redshift path
associated with metal lines and use all redshifts between $z_{\rm min}^{q}$ and
$z_{\rm max}^{q}$. This does not change the results significantly, but cuts down 
the computation time substantially. 
The results are listed in Table~\ref{table-jnu} and are plotted in 
Figure~\ref{fig:lowzcomp}.
For comparison, we also give the solutions for $J(\nu_{0})$ 
found using the standard parameters, ($\Omega_{\rm M}$,$\Omega_{\Lambda}$)~=~(1.0,0.0),
with this redshift path neglected. 
We find that ($\Omega_{\rm M}$,$\Omega_{\Lambda}$)~=~(0.3,0.7),
does not change the value of $J(\nu_{0})$ derived significantly from the
value found using ($\Omega_{\rm M}$,$\Omega_{\Lambda}$)~=~(1.0,0.0).

We performed a slightly modified re-analysis of the Scott et al.\ 
(2000b) sample of objects at $z \sim 2$ and
found little effect at high redshift as well.  
The solution found for ($\Omega_{\rm M}$,$\Omega_{\Lambda}$)~=~(1.0,0.0)
was log[$J(\nu_{0})$]$=-21.09^{+0.20}_{-0.17}$,
while for ($\Omega_{\rm M}$,$\Omega_{\Lambda}$)~=~(0.3,0.7), we find 
log[$J(\nu_{0})$]$=-21.25^{+0.20}_{-0.17}$ for these data.

\subsection{$d{\cal N}/dz$}
\label{sec-dndz}

In the case of a size distribution of Ly-$\alpha$ absorbers
that is constant in redshift, the evolution of the number
of Ly-$\alpha$ absorption lines per unit redshift is given by:
\begin{equation}
d{\cal N}/dz={\cal N}_{0} (1+z)^{2} [\Omega_{M} (1+z)^{3}  +
(1-\Omega_{M}-\Omega_{\Lambda}) (1+z)^{2} + \Omega_{\Lambda}]^{-0.5},
\label{equ:noevol}
\end{equation}
(Sargent et al.\ 1980) where ${\cal N}_{0}$ equals the
absorber cross section times the absorber comoving number density
times the Hubble distance, $\pi r_{0}^{2} \phi_{0} c H_{0}^{-1}$.
A plot of $d{\cal N}/dz$ versus $z$ for non-evolving Ly-$\alpha$ absorbers
in ($\Omega_{M}$,$\Omega_{\Lambda}$)~=~(1.0,0.0) and (0.3,0.7)
cosmologies is shown in Figure~\ref{fig:noevol}.
It is clear that non-evolving models are too shallow to fit points
at $z > 1.7$, so
the normalization is found from a fit to the FOS data.
The FOS data at $z < 1.7$ are consistent
with a non-evolving population for $(\Omega_{M},\Omega_{\Lambda})=(1.0,0.0)$.
The data are less consistent with a non-evolving concordance model in which
$(\Omega_{M},\Omega_{\Lambda})=(0.3,0.7)$, though not significantly so. 

The number density evolution of Ly-$\alpha$ absorbers over the 
redshift range $z=0-5$ cannot be approximated with a single power
law. There is a significant break in the slope of the
line number density with respect to redshift, near $z=1.7$ (Weymann et al.\ 1998, Paper IV)
though Kim, Cristiani, \& D'Odorico (2001) argue that the break occurs at
$z=1.2$.
Dav\'{e} et al.\ (1999) show from hydrodynamical simulations of the low
redshift Ly-$\alpha$ forest, that the evolution of the line density
is sensitive mainly to the HI photoionization rate, but also to the
evolution of structure (cf.\ their Figure 7).   The flattening of $d{\cal N}/dz$ observed by
Weymann et al.\ (1998) is mostly attributed to a dramatic decline in
$\Gamma(z)$ with decreasing $z$.
Dav\'{e} et al.\ (1999) derive an expression for the density of Ly-$\alpha$ forest
lines per unit redshift as a function of the HI photoionization rate:
\begin{equation}
\frac{d{\cal N}}{dz} = C [ (1+z)^{5} \Gamma^{-1}(z)]^{\beta-1} H^{-1}(z),
\label{equ:dave}
\end{equation}
where $C$ is the normalization at some fiducial redshift which we choose to be $z=0$
and $\Gamma (z)$ can be expressed by Equ.~\ref{equ:hmgam}.

We fit the FOS and MMT absorption line data,
binned in $d{\cal N}/dz$ as presented in Paper IV and
Scott et al.\ (2000a, Paper I), to this function
in order to derive the parameters describing $\Gamma (z)$ implied by the evolution
in Ly-$\alpha$ forest line density.
We observe flattening of $d{\cal N}/dz$ at $z < 1.7$, but
not to the degree seen by Weymann et al.\ (1998) in the Key Project data.
As described in Paper IV, we find $\gamma=0.54\pm0.21$, for 
lines above a 0.24~\AA\ threshold, while Weymann et al.\ (1998) measure 
$\gamma=0.15\pm0.23$.  See Paper IV for more discussion of the significance
and underlying causes of this difference.
We find $(A_{\rm HM},B_{\rm HM},z_{c},S)=(3.0\times 10^{-12},0.61,5.5 \times 10^{-7}, 7.07)$
and $(1.9 \times 10^{-11}, 0.38, 3.4 \times 10^{-7}, 6.21)$ for
$(\Omega_{\rm M},\Omega_{\Lambda})=(1.,0.)$ and
lines with rest equivalent widths above 0.24 and 0.32~\AA\, respectively.
These fits to Equ.~\ref{equ:dave} are shown in Figure~\ref{fig:dndz}(a).
In panel (b), we plot $\Gamma (z)$, as expressed in Equ.~\ref{equ:hmgam}, evaluated
using the parameters found from the fit to Equ.~\ref{equ:dave} above.
The HM96 solution and the
solution derived from the full FOS and MMT data sets are represented by the thick
and thin solid lines respectively.
The small values of $z_{c}$ derived from $d{\cal N}/dz$ above translate into
ionization rates that do not decrease dramatically with decreasing redshift
and result from the less pronounced flattening of $d{\cal N}/dz$ relative
to the Key Project.
These fits are particularly insensitive to the normalization, $A_{\rm HM}$, so
the errors on this parameter are large.  These fits should therefore not be
interpreted as measurements of $\Gamma (z)$ as reliable as those
found directly from the absorption line data.  But we find them instructive
nonetheless.
The observed $\Gamma (z)$ falls short of the ionization rate needed
to fully account for the change in the Ly-$\alpha$ line density with redshift,
indicating that if the value of $\gamma$ at low redshift is indeed slightly larger than that found 
by the Key Project, $d{\cal N}/dz$ may still be consistent with a 
non-evolving population of Ly-$\alpha$ absorbers in the sense noted above, but
the formation of structure in the low redshift universe
must play a significant role in determining the character of the Ly-$\alpha$ forest
line density.

\subsection{Comparison with Previous Results}
\label{sec-prevres}

\subsubsection{Proximity Effect}

KF93 performed a similar measurement with a small subsample of this
total sample- the HST Quasar Absorption Line Key Project
data of Bahcall et al.\ (1993).  We compare
our result to that from Sample 2 of KF93, which was constructed from the
Bahcall et al.\ (1993) data excluding one BAL quasar and all heavy
element absorption systems.
The Key Project sample has since been supplemented (Bahcall et al.\ 1996, Jannuzi et
al. 1998) and those data have been included when appropriate in the complete archival sample of FOS
spectra presented in Paper III.

The mean intensity KF93 derive from their Sample 2
($b=35$ km~s$^{-1}$, $\beta$=1.48, $\gamma$=0.21)
is $5.0^{+20.}_{-3.4} \times 10^{-24}$
ergs s$^{-1}$ cm$^{-2}$ Hz$^{-1}$ sr$^{-1}$.  
This result is lower than ours for $z < 1$ by a factor of $\sim 13$, though 
the errors are large on both results are large enough that they are 
consistent.  We use 162 lines in our low redshift solution for $J(\nu_{0})$,
65 more than KF93.

\subsubsection{Direct Measurements}
\label{sec-direct}

Several authors have examined the sharp cutoffs observed
in the HI disks of galaxies in the context of using these signatures to infer
the local ionizing background (Maloney 1993, Corbelli \& Salpeter 1993, Dove \& Shull 1994).
The truncations are modeled as arising primarily
from photoionization of the disk gas by the local extragalactic background radiation
field.  Using 21 cm observations (Corbelli, Scheider, \& Salpeter 1989, van Gorkom 1993)
to constrain these models, limits on the local ionizing background are placed at
$10^{4} < \Phi_{{\rm ion}} < 5 \times 10^{4}$ cm$^{-2}$ s$^{-1}$, where 
\begin{equation}
\Phi_{{\rm ion}} = 2\pi \int_{0}^{1} \mu d\mu \int_{\nu_{0}}^{\infty} \frac{J_{\nu}}{h\nu} d\nu =
\frac{\pi J(\nu_{0})}{h \alpha_{s}},
\end{equation}
and where $J_{\nu}=I_{\nu}$ for an isotropic radiation field.

Additionally, narrow-band and Fabry-Perot observations of
H$\alpha$ emission from intergalactic clouds (Stocke et al.\ 1991, Bland-Hawthorn et al.\ 1994,
Vogel et al.\ 1995, Donahue, Aldering, \& Stocke 1995)
place limits of $\Phi_{{\rm ion}} \lesssim 10^{4}$ cm$^{-2}$ s$^{-1}$, or
$J(\nu_{0}) < 7.6 \times 10^{-23}$ ergs s$^{-1}$ cm$^{-2}$ Hz$^{-1}$ sr$^{-1}$ for
$\alpha_{s}=1.8$, while
results from measurements of 
Galactic high velocity clouds
(Kutyrev \& Reynolds 1989, Songaila, Bryant, \& Cowie 1989,
Tufte, Reynolds, \& Haffner 1998)
imply $\Phi_{{\rm ion}} \lesssim 6 \times 10^{4}$ cm$^{-2}$ s$^{-1}$,
though the ionization of high velocity
clouds may be contaminated by a Galactic stellar contribution.  

Tumlinson et al.\ (1999) have reanalyzed the
3C273/NGC3067 field using the H$\alpha$ imaging data from Stocke et al.\ (1991) as well
as new GHRS spectra of 3C273, in order to model the ionization balance in 
the absorbing gas in the halo of NGC3067.  From this analysis, they derive the
limits, $2600 < \Phi_{{\rm ion}} < 10^{4}$ cm$^{-2}$ s$^{-1}$, 
or $10^{-23} < J(\nu_{0}) < 3.8 \times 10^{-23}$ ergs s$^{-1}$ cm$^{-2}$ Hz$^{-1}$ sr$^{-1}$
at $z = 0.0047$.
Weymann et al.\ (2001) have recently reported an upper limit of  
$\Phi_{{\rm ion}} < 1.01 \times 10^{4}$ cm$^{-2}$ s$^{-1}$, or
$J(\nu_{0}) < 3.84 \times 10^{-23}$ ergs s$^{-1}$ cm$^{-2}$ Hz$^{-1}$ sr$^{-1}$ from 
Fabry-Perot observations of the intergalactic HI cloud, 1225+01, for a face-on
disk geometry.  If an inclined disk geometry is assumed, this lower limit becomes
$J(\nu_{0}) < 9.6 \times 10^{-24}$ ergs s$^{-1}$ cm$^{-2}$ Hz$^{-1}$ sr$^{-1}$.
These results are summarized in Figure~\ref{fig:allzcomp}.
It is
encouraging that the proximity effect value is consistent with the limits on
the background set by these more direct estimates which are possible locally.

\subsection{Comparison with Models}
\label{sec-models}

Haardt \& Madau (1996) calculated the spectrum of the UV background as
a function of frequency and redshift using a model based on the integrated emission from
QSOs alone.  The QSO luminosity function is drawn from Pei (1995).
The opacity of the intergalactic medium is computed from the observed redshift and
column density distributions of Ly-$\alpha$ absorbers given by Equ.~\ref{eq:dndzdnh1}. 
The effects of attenuation and reemission of radiation 
by hydrogen and helium in Ly-$\alpha$ absorbers are included in these models.
Their result for $q_{0}=0.5$ and $\alpha_{s}=1.8$ at $z=0$ is 
$J(\nu_{0}) = 1.6 \times 10^{-23}$ ergs s$^{-1}$ cm$^{-2}$ Hz$^{-1}$ sr$^{-1}$.  

Fardal et al.\ (1998)
compute opacity models for the intergalactic medium (IGM) based on 
high resolution observations of the high redshift Ly-$\alpha$ forest from
several authors. 
Shull et al.\ (1999) extend the models of Fardal et al.\ (1998) to $z=0$,
treating opacity of low redshift 
Ly-$\alpha$ forest from observations
made with HST/GHRS (Penton et al.\ 2000a,b) and with HST/FOS (Weymann et al.\ 1998). 
Like Haardt \& Madau (1996), they also incorporate the 
observed redshift distribution of Lyman limit systems
with log(N$_{{\rm HI}}$) $ > 17$ (Stengler-Larrea et al.\ 1995, Storrie-Lombardi et al.\ 1994).
Their models also allow for a contribution from star formation in galaxies in addition to AGN.
The QSO luminosity function again is taken to follow the
form given by Pei (1995) with upper/lower cutoffs at 0.01/10 L$_{*}$.
QSO UV spectral indicies are assumed to equal 0.86, while the
ionizing spectrum at $\nu > \nu_{0}$ has
$\alpha_{s}=1.8$.  
The contribution to the background from stars was normalized to the
H$\alpha$ luminosity function observed by Gallego et al.\ (1995) and
the escape fraction of photons of all energies from galaxies was taken to be
$<f_{{\rm esc}}> =0.05$. 
The full radiative transfer model described in Fardal et al.\ (1998) 
was used to calculate the contribution to the mean intensity by AGN, but not the
contribution from stars, as they were assumed to contribute no flux above 4 Ryd,
the energies at which the effects of IGM reprocessing become important.
These authors find $J(\nu_{0})=2.4 \times 10^{-23}$ ergs s$^{-1}$ cm$^{-2}$ Hz$^{-1}$ sr$^{-1}$
at $z\sim 0$, with approximately equal contributions
from AGN and stars, a value somewhat lower than our result for $z < 1$, but which is
allowed within the errors.

We estimate the contribution to the UV background from star-forming
galaxies using the galaxy luminosity function of the Canada-France Redshift Survey (Lilly
et al.\ 1995).  At $z \sim 0.5$, we derive $J^{\rm gal}(\nu_{0})= 1.5 \times 10^{-22}$
ergs s$^{-1}$ cm$^{-2}$ Hz$^{-1}$ sr$^{-1}$, assuming $<f_{{\rm esc}}>=1$.  The HM96
models for the QSO contribution give $J^{\rm QSO}(\nu_{0}) = 5.2 \times 10^{-23}$
ergs s$^{-1}$ cm$^{-2}$ Hz$^{-1}$ sr$^{-1}$ at $z \sim 0.5$.  These estimates, and the
range of measured $J(\nu_{0})$ in this paper, $\sim 5-16 \times 10^{-23}$
ergs s$^{-1}$ cm$^{-2}$ Hz$^{-1}$ sr$^{-1}$ imply an escape fraction of UV photons from
galaxies between 4\% and 70\%.
The $J(\nu_{0})$ inferred from $d{\cal N}/dz$ in Section~\ref{sec-dndz} implies
escape fractions well over 100\%. 

Bianchi et al.\ (2001) make updated estimates of the mean intensity of the background with
contributions from both QSOs and star-forming galaxies.  
Their models incorporate various values of the escape fraction of Lyman continuum
photons from galaxies which are constant with redshift and wavelength.
Our new results
at $z < 1.7$ are most consistent with their models of the QSO contribution alone,
though some contribution from galaxies, ie. a small f$_{\rm esc}$, 
is allowed within the uncertainties.
At $z \sim 3.5$, recent results from Steidel, Pettini, \& Adelberger (2001)
on the Lyman-continuum radiation from high redshift galaxies
suggest that these sources become a more important component of the UV background
at high redshift.

\subsection{Systematics}
\label{sec-systematics}

Drawing on lessons learned from our work on high redshift objects in Paper II, 
we have made corrections for quasar systemic redshifts before performing the
proximity effect analysis, as discussed in \S\ref{sec-zsys}.
This correction, $\sim 300$ km~s$^{-1}$,  
was made to QSO redshifts measured from
Ly-$\alpha$ emission for objects for which no systemic redshift measurement was
available.  For the low redshifts considered in this paper, redshifts
measured from [OIII], MgII, or Balmer emission lines were deemed suitable as
QSO systemic redshift measurements.

We have removed known gravitational lenses from the sample.  As discussed above, 
we perform the proximity effect analysis omitting and including
spectra that show associated absorption and damped Ly-$\alpha$ absorption
and determined that neither of these populations significantly biases our results.

Because we are working with low redshift data where line densities are low, 
we expect that blending has not contributed as strong a systematic
effect as in the high redshift sample of Paper II.  The curve-of-growth effects
discussed in Paper II may still be present, since many lines in the
sample have equivalent widths which place them on the flat part of the
curve of growth.

However, the effects of clustering may be even more important at low
redshift than at high redshift.
Loeb \& Eisenstein (1995) showed how the fact that quasars reside in
the dark matter potentials of galaxies and small groups of galaxies
can influence the proximity effect signature.   The peculiar velocities of
matter clustered 
in these potentials can result in Ly-$\alpha$ absorption at redshifts
greater than the quasar emission redshift.  
We found that including associated absorbers in our sample did not
significantly change our results.  Recently, Pascarelle et al.\ (2001)
report evidence for a lower incidence of Ly-$\alpha$ absorption lines arising in 
the gaseous halos of galaxies in the vicinities
of QSOs than in regions far from QSOs.  They argue that galaxy-QSO clustering
may lead proximity effect measurements to overestimate $J(\nu_{0})$ at $z < 1$ by
a up to a factor of 20.  While we agree that most
systematic effects in this type of analysis, including clustering, will lead to 
overestimates of $J(\nu_{0})$, the agreement between our results and the direct
measurements discussed in Section~\ref{sec-direct} give us confidence that our
results are not biased by this large a factor.

The hydrodynamic simulations
of the low redshift Ly-$\alpha$ forest of Dav\'{e} et al.\ (1999) indicate that, at
low redshift, structures of the same column density correspond to
larger overdensities and more advanced dynamical states than at high redshift.  
For a $(\Omega_{\rm M},\Omega_{\Lambda})~=~(0.4,0.6)$ cosmology,
an equivalent width limit of 0.32~\AA\ corresponds to an overdensity
of $\sim 1.4$ at $z \sim 3$, while at $z \sim 0.6$, this limit corresponds to 
$\rho_{H} / \overline{\rho_{H}} \sim 13$.   This may have implications on the clustering of Ly-$\alpha$ 
absorption lines around QSOs and hence on the values of 
$J(\nu_{0})$ derived from the proximity effect.
It is possible that we are seeing this clustering effect in the
variable threshold solution at $z > 1$, in which the two
highest $\omega(z)$ lines in the sample
are responsible for the inability to isolate a reasonable 
maximum likelihood $J(\nu_{0})$.

\section{Summary}
\label{sec-summary}

We have analyzed a set of 151 QSOs and 906 Ly-$\alpha$ absorption lines, 
the subset of the total data set presented in Paper III that is
appropriate for the proximity effect.
The primary results of this paper are as follows:

(1) At low redshift, Balmer, [OIII], and Mg II emission lines are
reasonable indicators of QSO systemic redshifts.  Ly-$\alpha$ emission
is blueshifted by $\sim 300$ km~s$^{-1}$ with respect to [OIII].

(2) The value of $J(\nu_{0})$ is observed to increase with redshift
over the redshift range of the sample data, $0.03 < z < 1.67$. 
Dividing the sample at $z = 1$, we find $J(\nu_{0})=
6.5^{+38.}_{-1.6} \times 10^{-23}$ ergs s$^{-1}$ cm$^{-2}$ Hz$^{-1}$ sr$^{-1}$,
at low redshift and $J(\nu_{0})=
1.0^{+3.8}_{-0.2} \times 10^{-22}$ ergs s$^{-1}$ cm$^{-2}$ Hz$^{-1}$ sr$^{-1}$
at high redshift.  

(3)  The inclusion of blazars at $z < 1$ has no significant effect on the result.
There is no significant difference between the values of $J(\nu_{0})$  
derived from radio loud (RL $>$ 1.0) and radio quiet (RL $<$ 1.0) objects,
indicating that the observed richness of quasar environments does not
distinctly bias the proximity effect analysis. 

(4) Using information measured and gathered from the literature on each QSO's UV spectral 
index and solving for the HI ionization rate, yields  
$1.9 \times 10^{-13}$ s$^{-1}$ for $z < 1$ and
$1.3 \times 10^{-12}$ s$^{-1}$ for and $z > 1$.  
Solving directly for
the parameters $(A_{\rm HM},B_{\rm HM},z_{c},S)$ in the HM96 parametrization of $\Gamma(z)$
using the HST/FOS data presented by Bechtold et al.\ (2001) combined with the
high redshift, ground-based data presented by Scott et al.\ (2000a,b) results
in $(A_{\rm HM},B_{\rm HM},z_{c},S)$~=~($7.6\times 10^{-13}$, 0.35, 2.07, 1.77) for $\beta=1.46$ and
$(A_{\rm HM},B_{\rm HM},z_{c},S)=(3.2\times 10^{-13}$, 1.45, 2.13, 1.42) for 
$\beta=1.7$ for $1.7 < z < 3.8$.

(5) Allowing for a varying equivalent width threshold across each QSO spectrum results
in consistently higher values of $J(\nu_{0})$ than are found from the constant threshold treatments.  
At $z > 1$, the variable threshold solution is not
well-constrained. Jackknife experiments indicate that this is due the objects 0743-6719 and
0302-2223,
namely the highest $\omega(z)$ absorption lines in each of their spectra.

(6) Allowing for a cosmology in which
$(\Omega_{\rm M},\Omega_{\Lambda}) = (0.3,0.7)$, rather
than (1.,0.) has no significant effect on the value of $J(\nu_{0})$
derived from these data.

(7) The $z < 1$  result is in agreement with the range of values of 
the mean intensity of the hydrogen-ionizing background allowed by
a variety of local estimates, including H$\alpha$ imaging and modeling of
galaxy HI disk truncations.  To within the uncertainty in the measurement,
this result agrees with the one previous proximity effect
measurement of the low redshift UV background (KF93).  These results are
consistent with
calculated models based upon the integrated emission from QSOs alone (HM96)
and with models which include both QSOs and starburst galaxies (Shull et al.\ 1999).
The uncertainties do not make a distinction between these two models possible.

(8) The results presented here tentatively confirm the IGM evolution scenario
provided by large scale hydrodynamic simulations (Dav\'{e} et al.\ 1999).  This
scenario, which is successful in describing many observed properties of the low redshift
IGM, is dependent upon an evolving $J(\nu_{0})$ which decreases from $z = 2$ to
$z = 0$.   However, the low redshift UV background required to match
the observations of the evolution of the Ly-$\alpha$ forest line density
is larger than found from the data, indicating that structure formation is
playing a role in this evolution as well.
Our results and the work of others are summarized in Figure~\ref{fig:allzcomp}.
We find some evidence of evolution in $J(\nu_{0})$, though it appears that even larger
data sets, especially at $z < 1$ and/or improved proximity effect ionization models 
will be required to improve the significance.

\acknowledgements
The authors thank the anonymous referee for a careful review of the paper
and for helpful suggestions.
This research has made use of the NASA/IPAC Extragalactic Database (NED)
which is operated by the Jet Propulsion Laboratory, California Institute
of Technology, under contract with the National Aeronautics and Space
Administration.
This project was supported by STScI grant No.\
AR-05785.02-94A and STScI grant No.\ GO 066060195A.
JS acknowledges support of the National Science Foundation
Graduate Research Fellowship and the Zonta Foundation Amelia Earhart
Fellowship.  JS, JB, and MM received financial support
from NSF grant AST-9617060B.
AD acknowledges support from NASA Contract No.\
NAS8-39073 (CXC).
VPK acknowledges partial support from an award from the William F. Lucas
Foundation and the San Diego Astronomers' Association.

\clearpage

\clearpage

\clearpage

\begin{figure}
\plotone{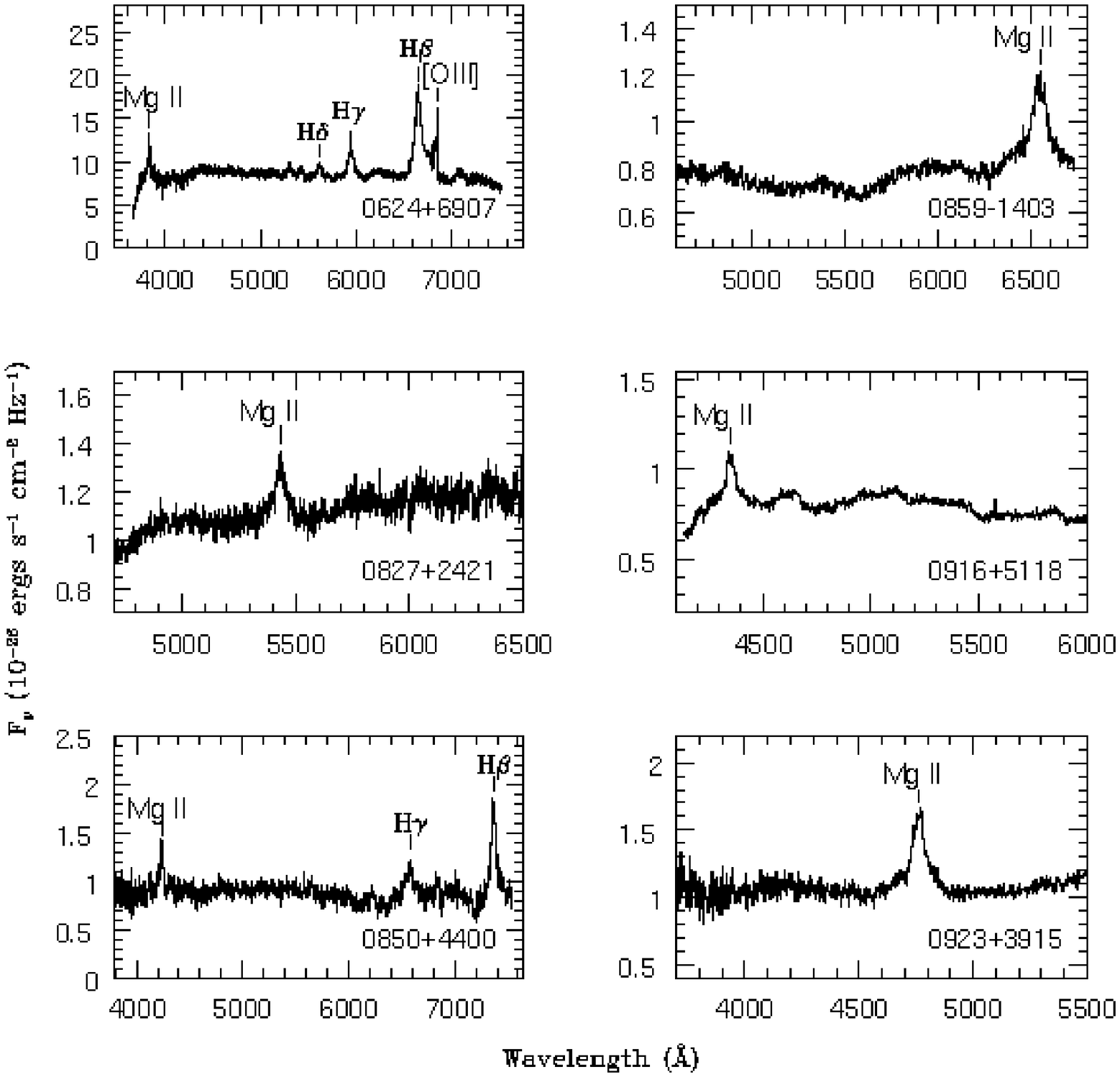}
\end{figure}

\clearpage

\begin{figure}
\plotone{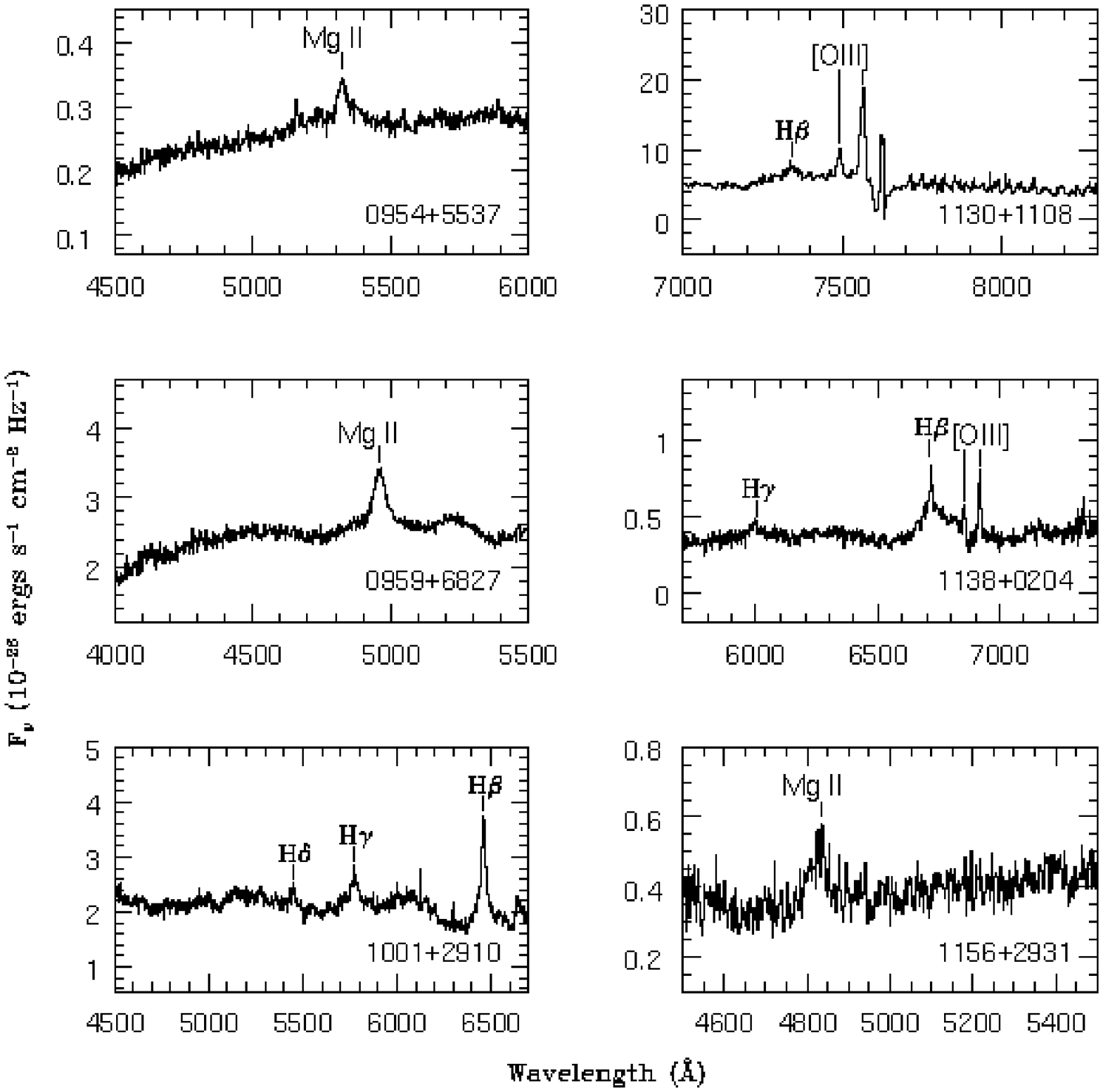}
\end{figure}

\clearpage

\begin{figure}
\plotone{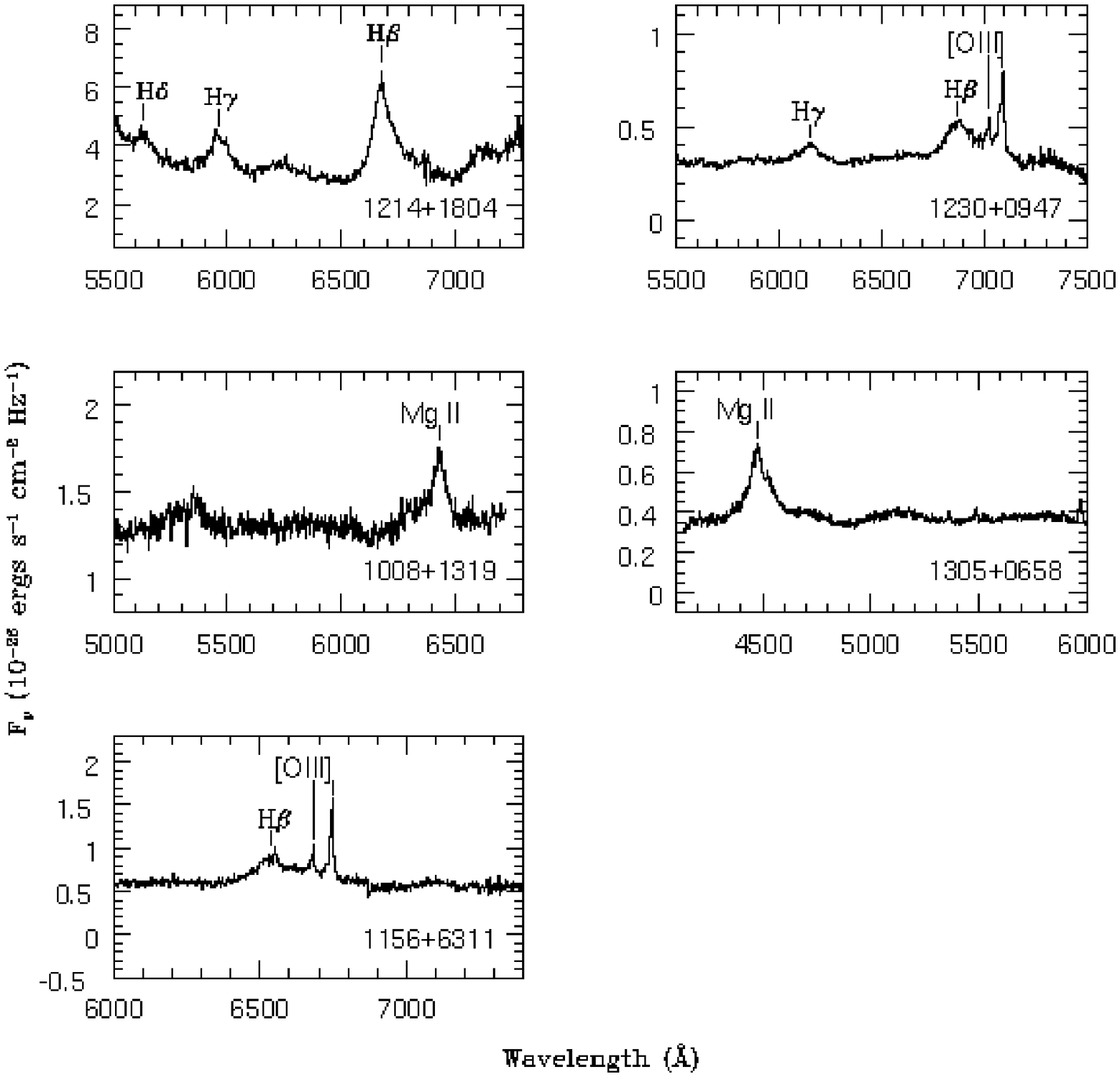}
\end{figure}

\clearpage

\begin{figure}
\plotone{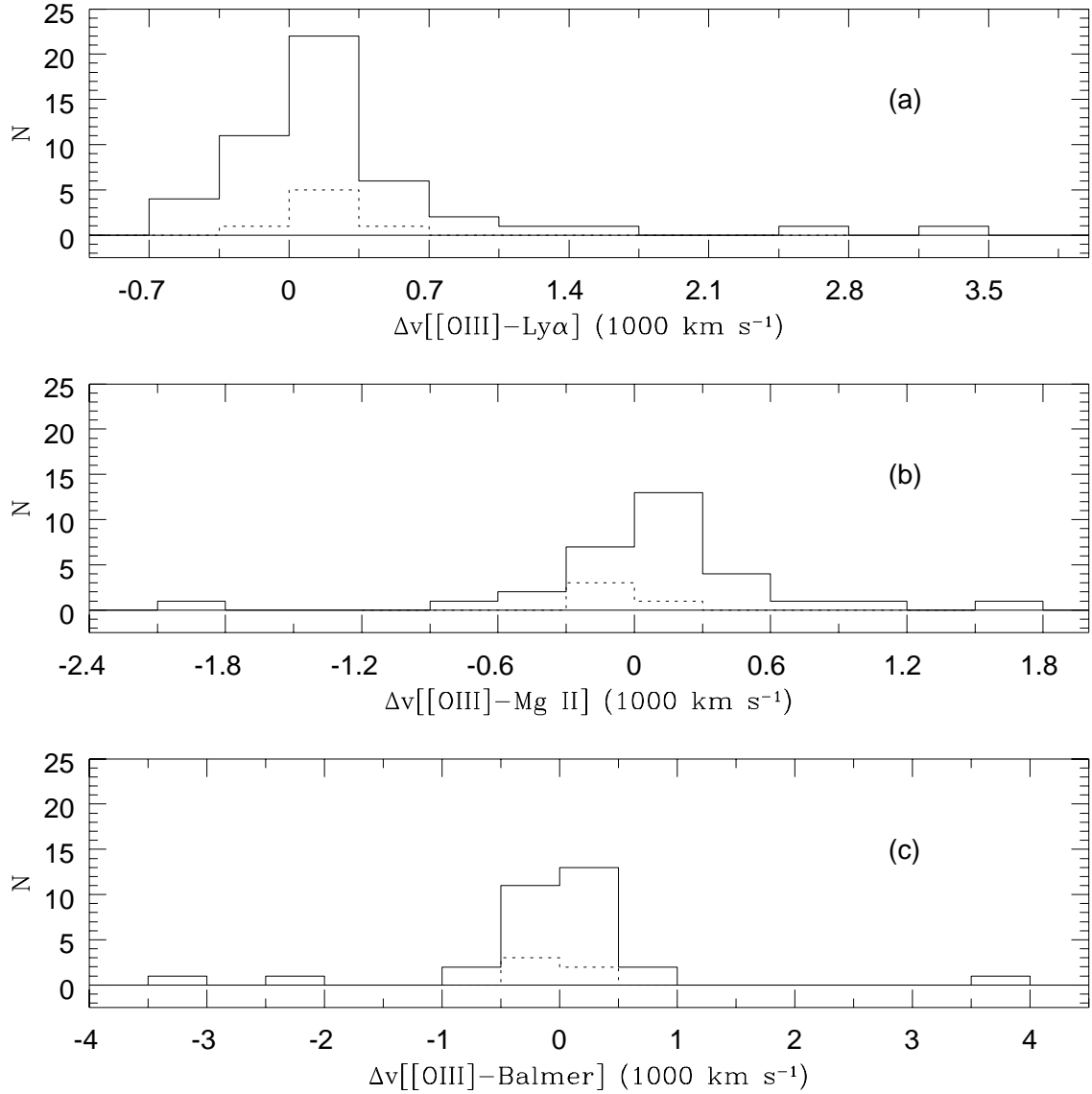}
\caption
{Histograms of redshift differences between [OIII] and
(a) Ly-$\alpha$, (b) Mg II, and (c) Balmer emission lines, dotted
lines show results from Laor et al.\ (1995)
\label{fig-zhist} }
\end{figure}

\clearpage

\begin{figure}
\plotone{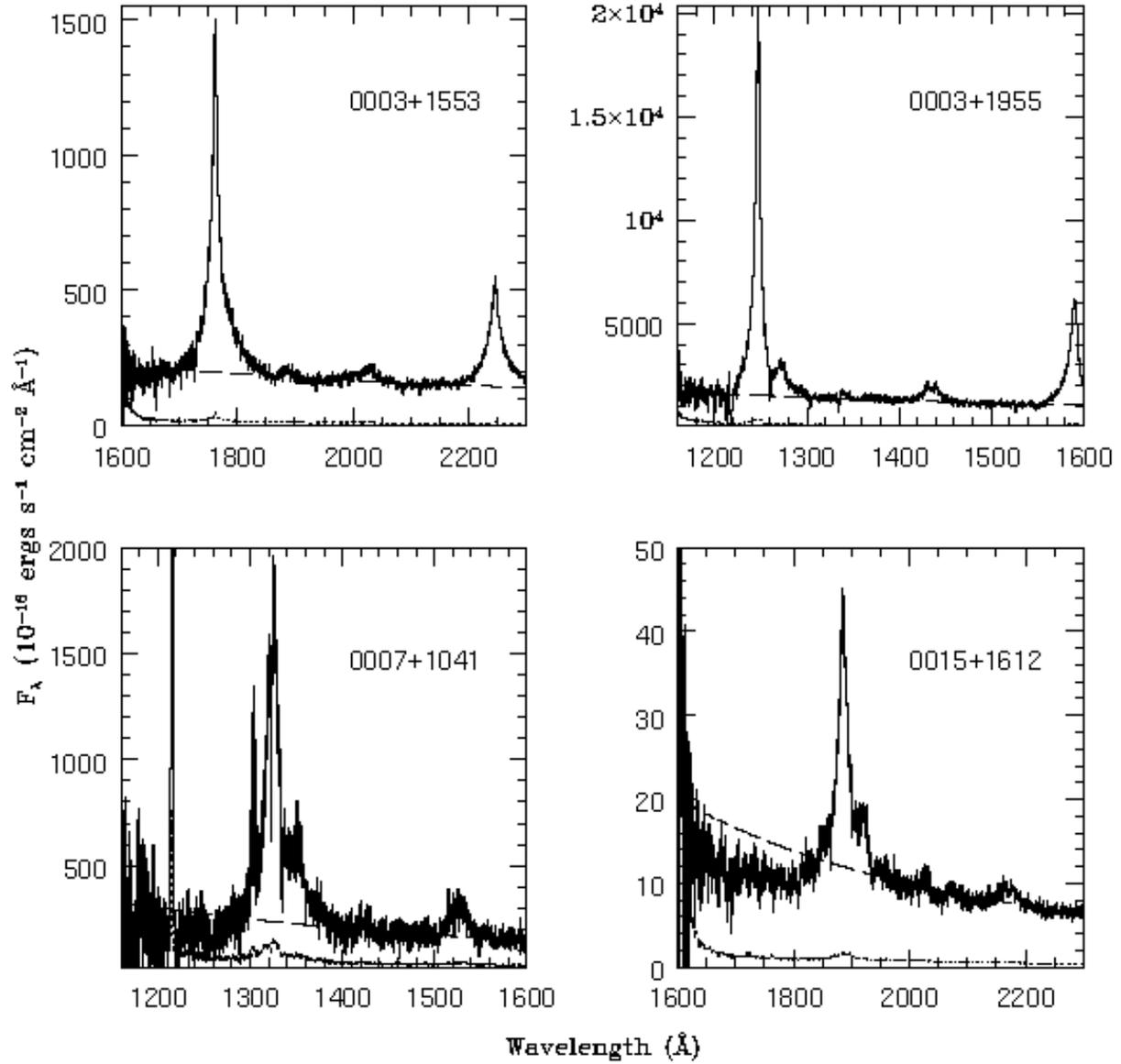}
\caption
{FOS spectra used to extrapolate to the Lyman limit flux of each object:
(dashed lines) power law fits, (dotted lines) 1$\sigma$ errors in spectrum \label{fig-fnus}}
\end{figure}

\clearpage

\begin{figure}
\noindent Remainder of Figure 4 available at:

\vspace{0.3in}
{\tt http://lithops.as.arizona.edu/$\tilde{\;}$jill/QuasarSpectra/}

\vspace{0.3in}
\noindent or

\vspace{0.3in}
{\tt http://hea-www.harvard.edu/QEDT/QuasarSpectra/}
\vspace{4.in}
\end{figure}

\clearpage

\begin{figure}
\plotone{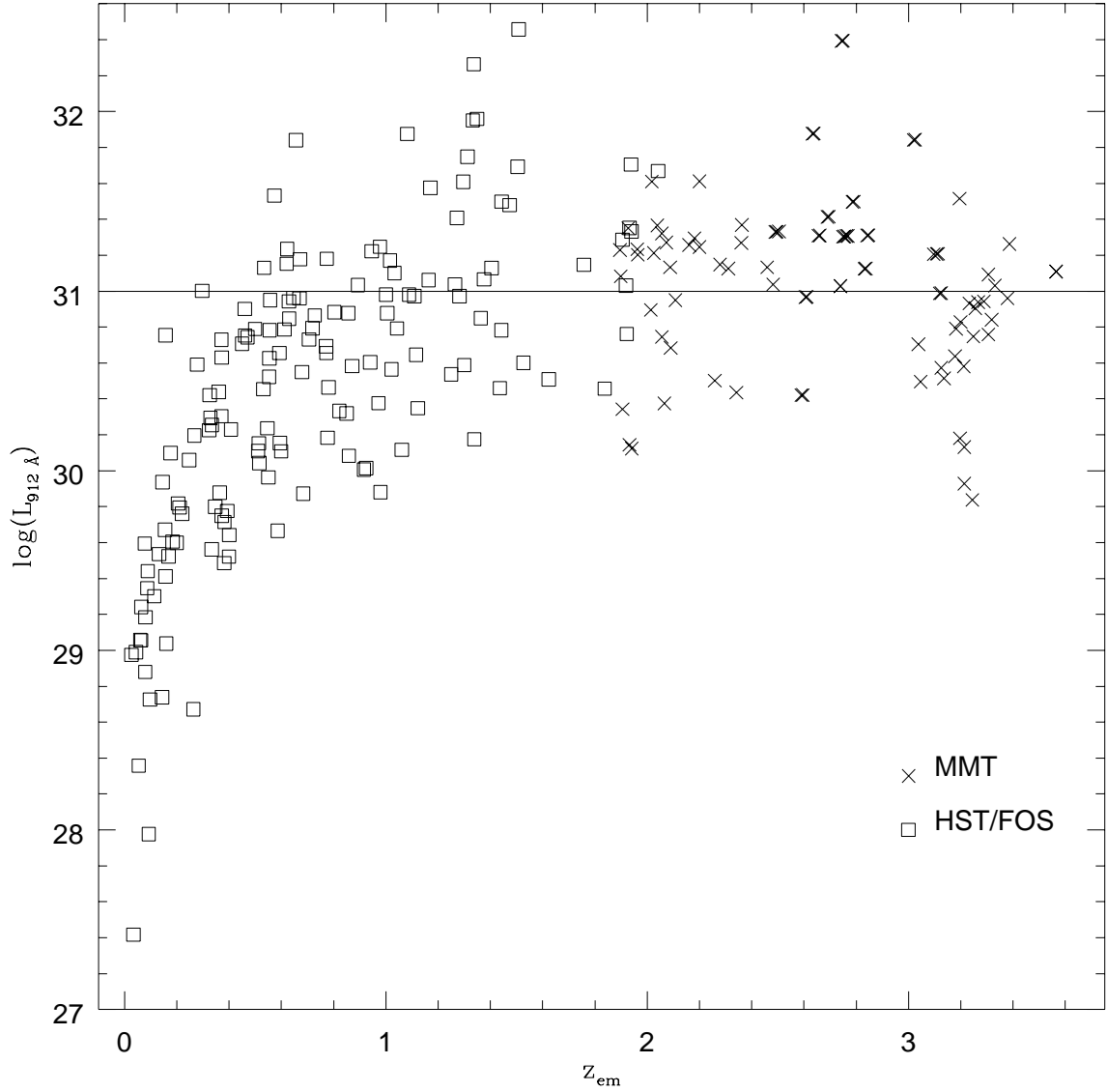}
\caption
{Lyman limit luminosity versus redshift for objects in the HST/FOS sample (squares) 
and in the MMT sample presented in Papers I and II (crosses), solid line
indicates the boundary between low and high luminosity objects \label{fig-zL}}
\end{figure}

\clearpage

\begin{figure}
\plotone{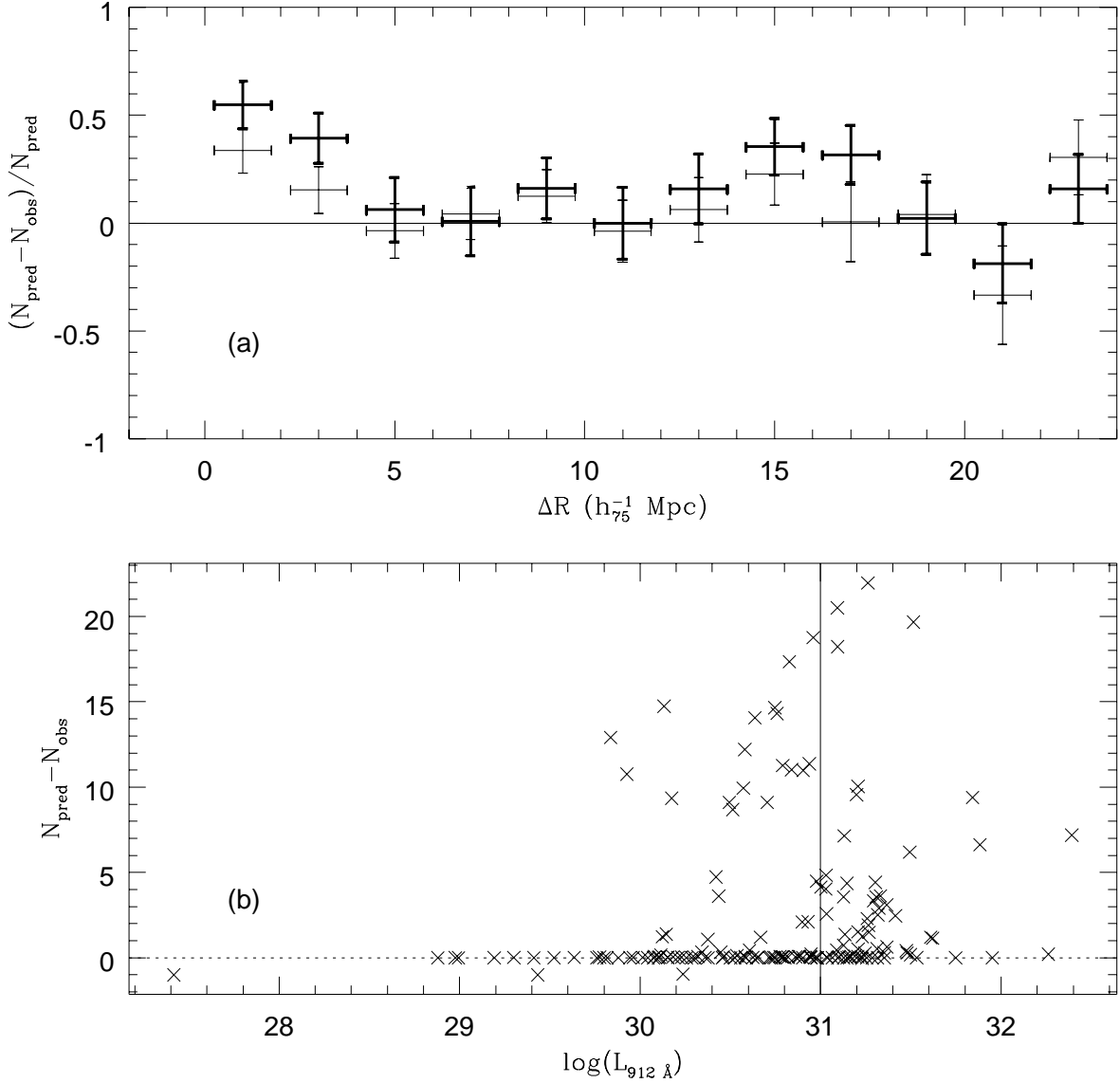}
\caption
{(a) Relative deficit of lines with respect to the number predicted by
Equ.~\ref{eq:dndzdw} for $W_{thr}= 0.32$~\AA\
versus distance from the QSO for high and low luminosity QSOs (thick and thin solid
lines, respectively) in both the HST/FOS sample presented
in Paper III and the MMT sample presented in Paper I;  
(b) Deficit of lines within 2 h$_{75}^{-1}$ Mpc as a function of QSO
Lyman limit luminosity for the HST/FOS and MMT samples, the vertical
line delineates the boundary between low and high luminosity objects
\label{fig-npred}}
\end{figure}

\clearpage

\begin{figure}
\plotone{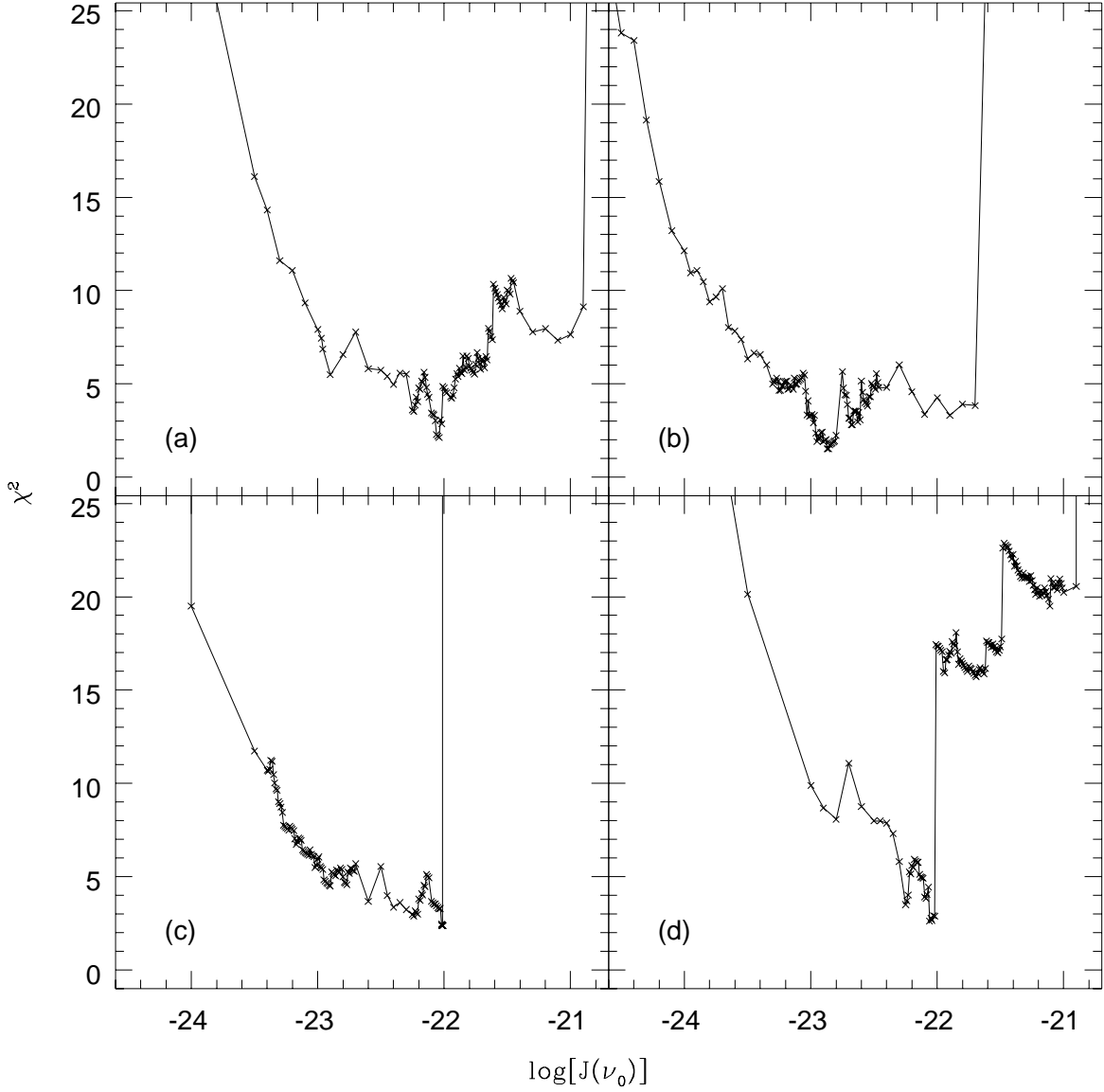}
\caption{
$\chi^{2}$ of binned data with respect to the ionization model
expressed in Equ.~\ref{equ:dndx2}
versus log[$J(\nu_{0})$] for various
redshift ranges and equivalent width thresholds: 
(a) $W_{thr}= 0.32$~\AA; 
(b) $W_{thr}= 0.32$~\AA, $z < 1$; 
(c) $W_{thr}= 0.32$~\AA, $z > 1$;
(d) $W_{thr}= 0.24$~\AA
\label{fig:chi2} }
\end{figure}

\clearpage

\begin{figure}
\plotone{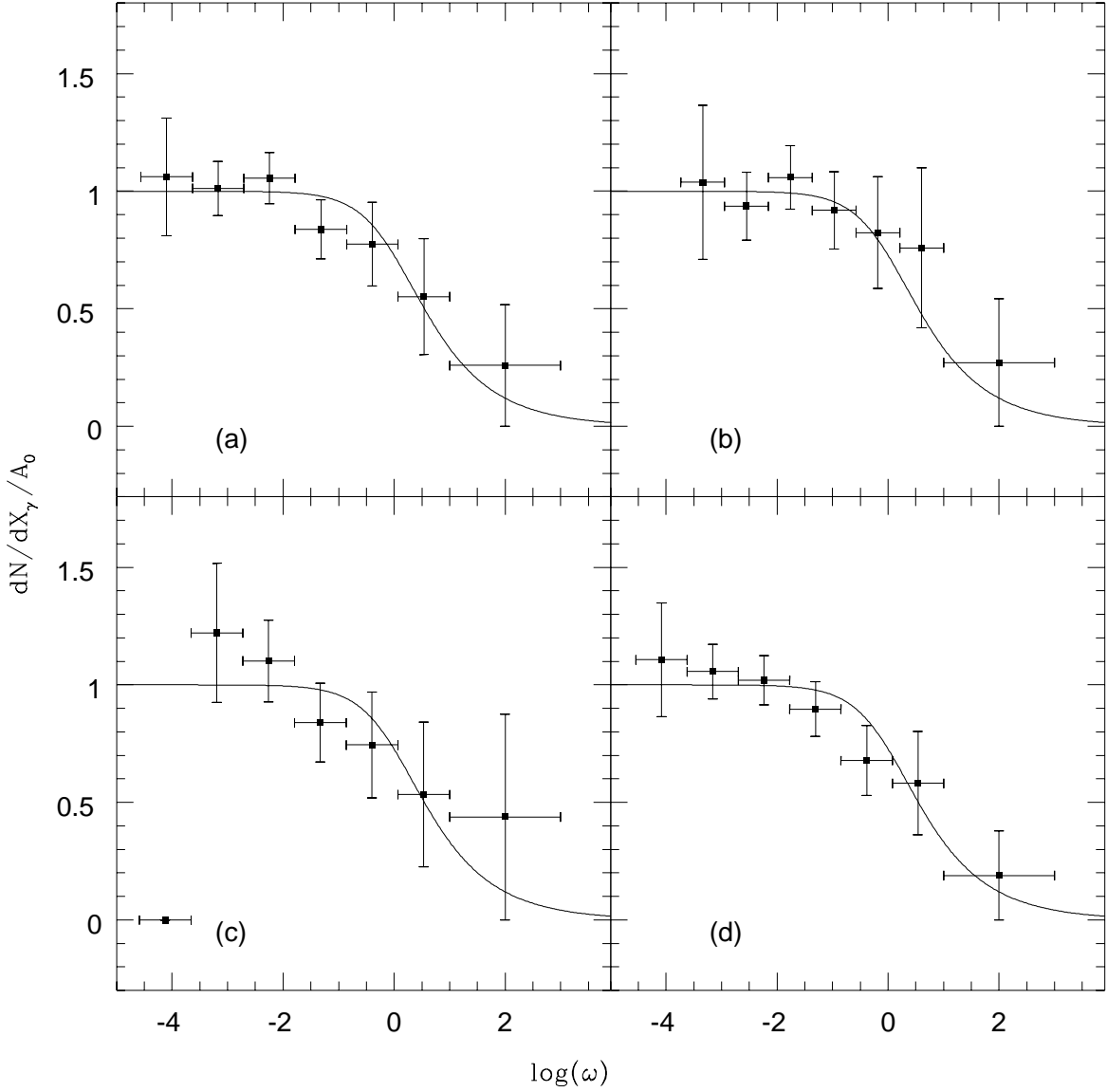}
\caption{
Number distribution per coevolving redshift coordinate expressed
in Equ.~\ref{equ:dndx2} for the best
fit values of $J(\nu_{0})$ (BDO method); (a-d) same as Fig.~\ref{fig:chi2}
\label{fig:dndx} }
\end{figure}

\clearpage

\begin{figure}
\plotone{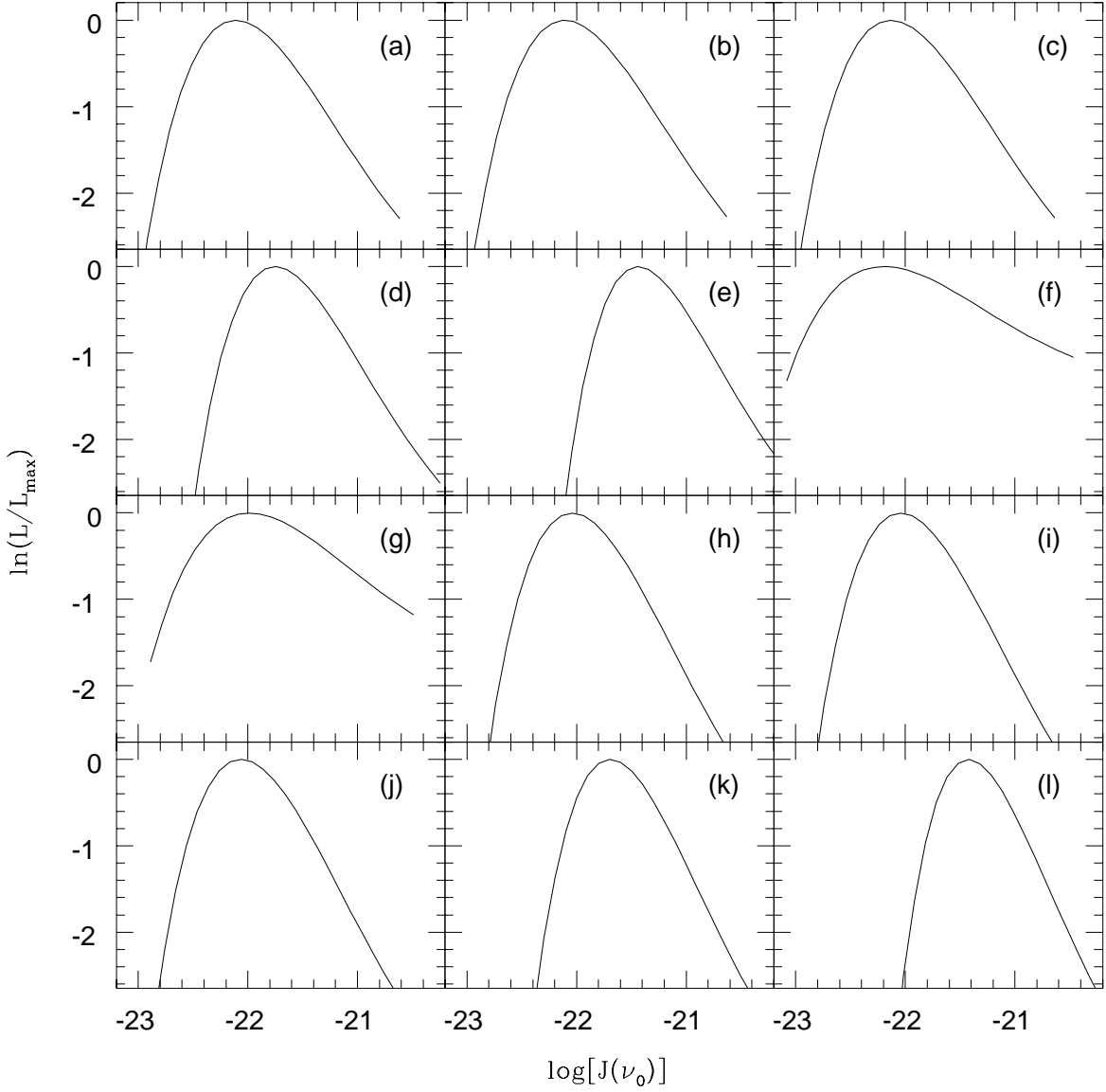}
\caption{
Likelihood function versus log[$J(\nu_{0})$] for $W_{thr}= 0.32$~\AA\, $(\beta,b)=$
(a) (1.46,35);
(b) (1.46,25);
(c) (1.45,25);
(d) (1.70,30);
(e) (2.04,25);
(f) (1.46,35), $z < 1$;
(g) (1.46,35), $z > 1$; 
and for $W_{thr}= 0.24$~\AA\, $(\beta,b)=$
(h) (1.46,35);
(i) (1.46,25);
(j) (1.45,25);
(k) (1.70,30);
(l) (2.04,25)
\label{fig:like1} }
\end{figure}

\clearpage

\begin{figure}
\plotone{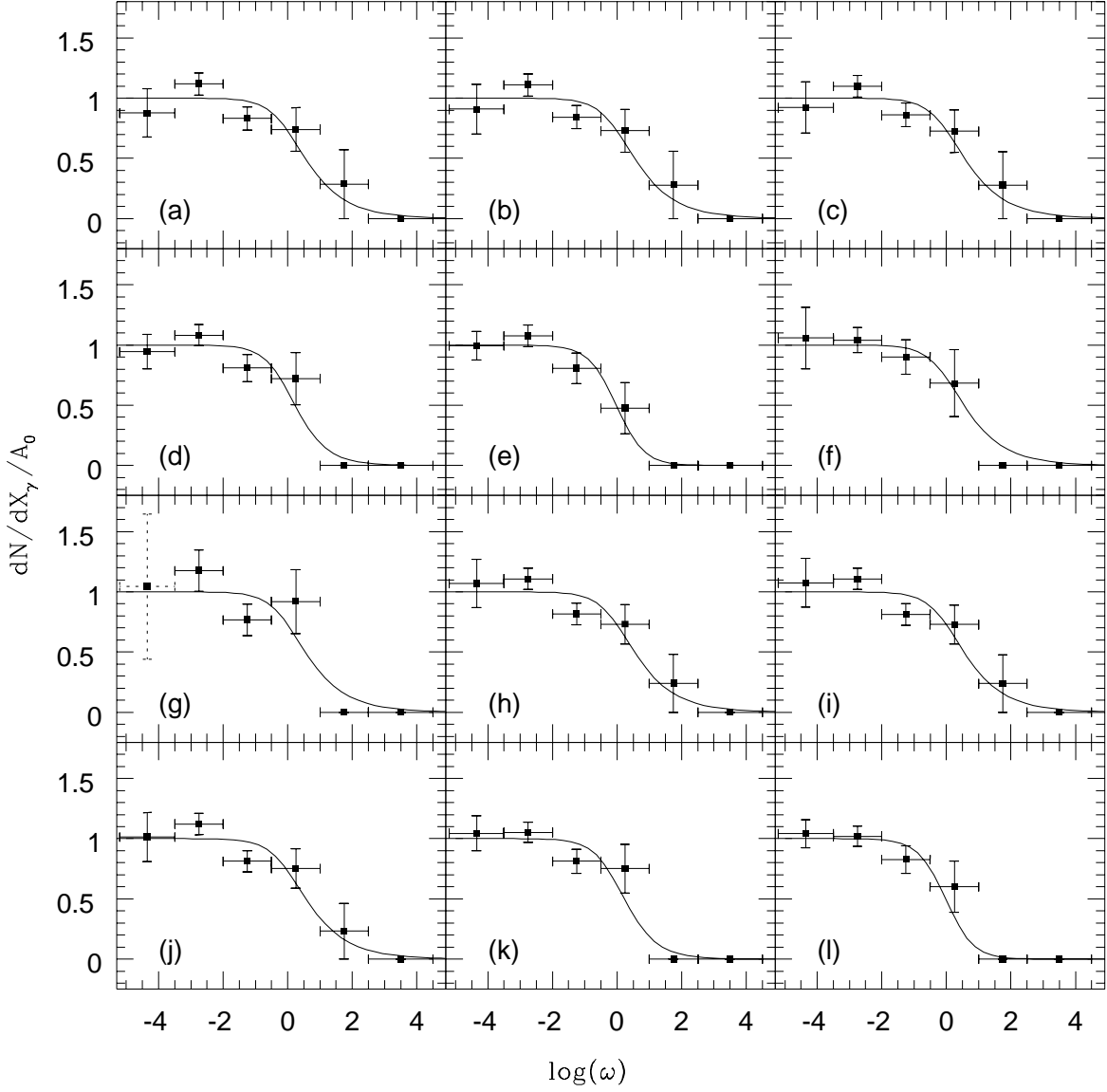}
\caption{
Number distribution per coevolving redshift coordinate for the best
fit values of $J(\nu_{0})$ (KF method); (a-l) same as Fig.~\ref{fig:like1};
the dotted point and error bars in (g) has been divided by 5 for clarity 
\label{fig:dndxl1} }
\end{figure}

\clearpage

\begin{figure}
\plotone{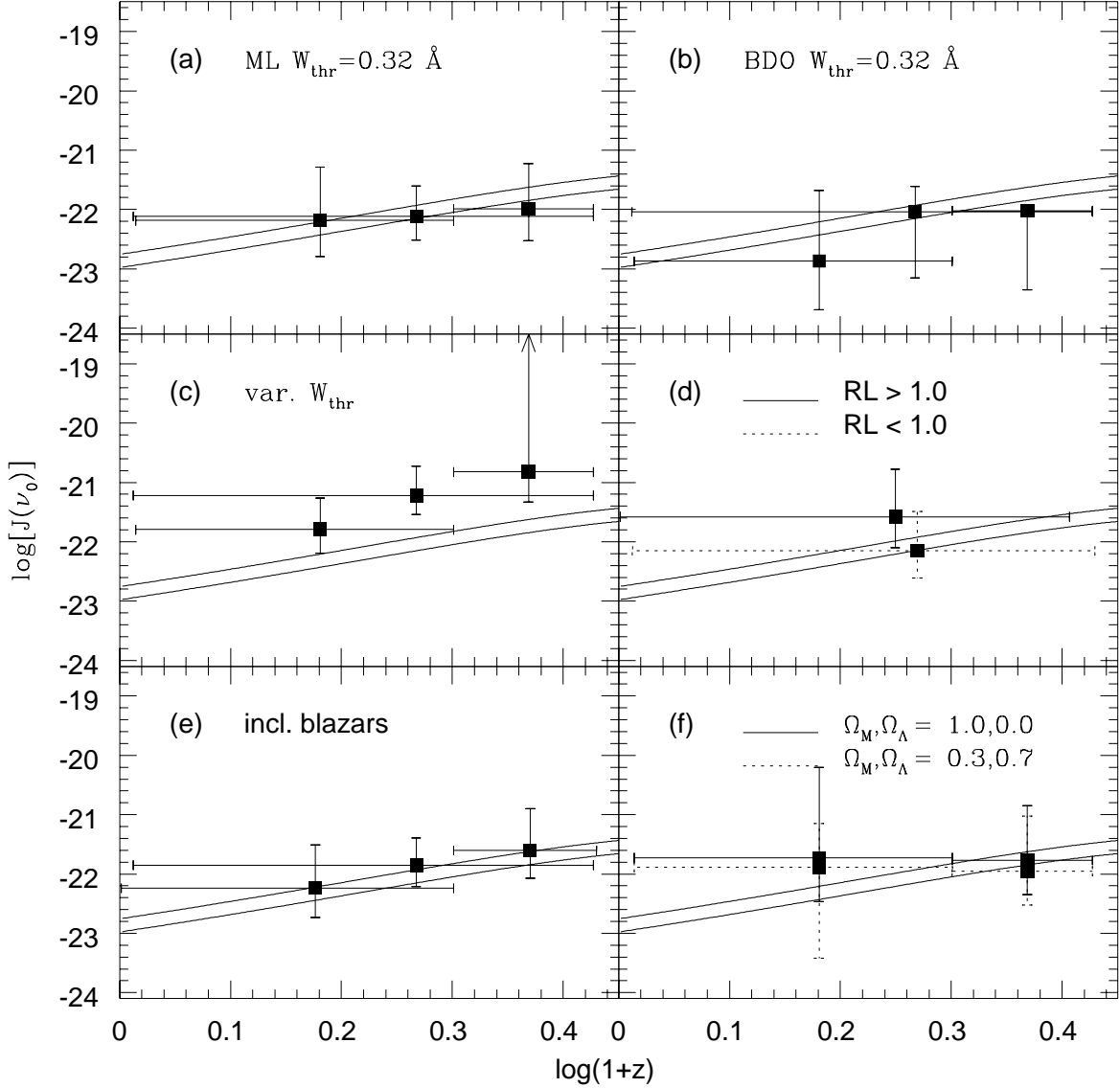}
\caption{
log[$J(\nu_{0})$] versus redshift, solid curves in (a)-(f) correspond
to HM96 models:  
(a) $W_{thr}=0.32$~\AA\, all redshifts, $0.03 < z < 1.67$,
and $z < 1$, $z > 1$ separately, ML method;
(b) same as (a), BDO method; 
(c) variable threshold, all redshifts $0.03 < z < 1.67$,
and $z < 1$, $z > 1$ separately; 
(d) $W_{thr}=0.32$~\AA\, all redshifts,
RL $>$ 0.3 and RL $<$ 0.3; 
(e) $W_{thr}=0.32$~\AA\, all redshifts,
$0.03 < z < 1.67$, and $z < 1$, total sample including blazars; 
(f) $W_{thr}=0.32$~\AA\, $z < 1$ and $z > 1$, 
(solid points) $(\Omega_{M},\Omega_{\Lambda})=(1.0,0.0)$,
(dotted points) $(\Omega_{M},\Omega_{\Lambda})=(0.3,0.7)$, 
metal line dz neglected in both cases
\label{fig:lowzcomp} }
\end{figure}

\clearpage

\begin{figure}
\plotone{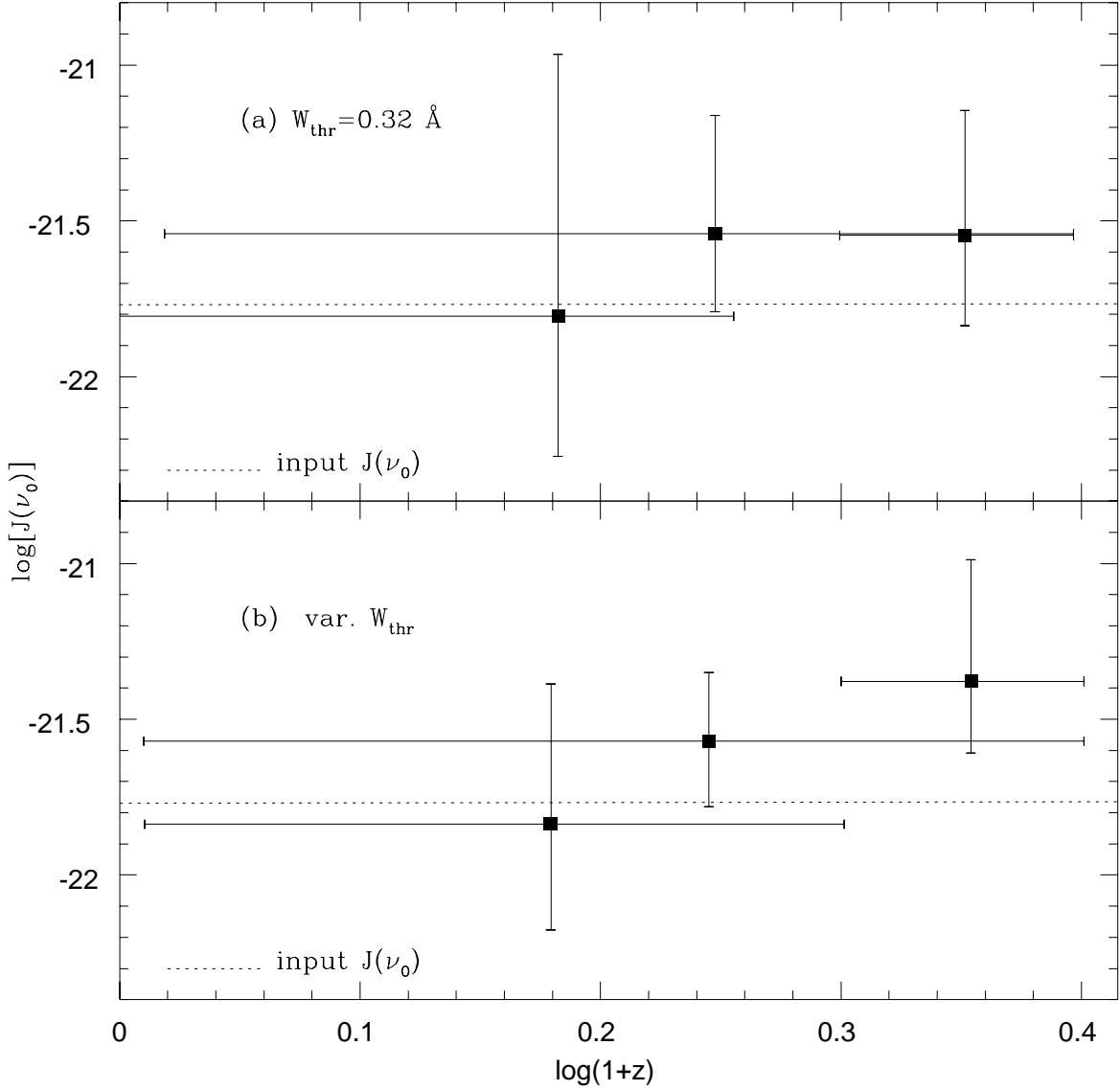}
\caption{
(a) Values of log[$J(\nu_{0})$] recovered from simulated QSO spectra with
proximity effect included:
(dotted lines)- input $J(\nu_{0},z)$,
see Figure~\ref{fig:lowzcomp}(a)
(solid points)- recovered $J(\nu_{0})$ for
$W_{thr}=0.32$~\AA\ at all redshifts and at $z < 1$ and $z > 1$ separately;
(b) same as (a), but $J(\nu_{0})$ recovered using variable threshold
\label{fig:sims} }
\end{figure}

\clearpage

\begin{figure}
\plotone{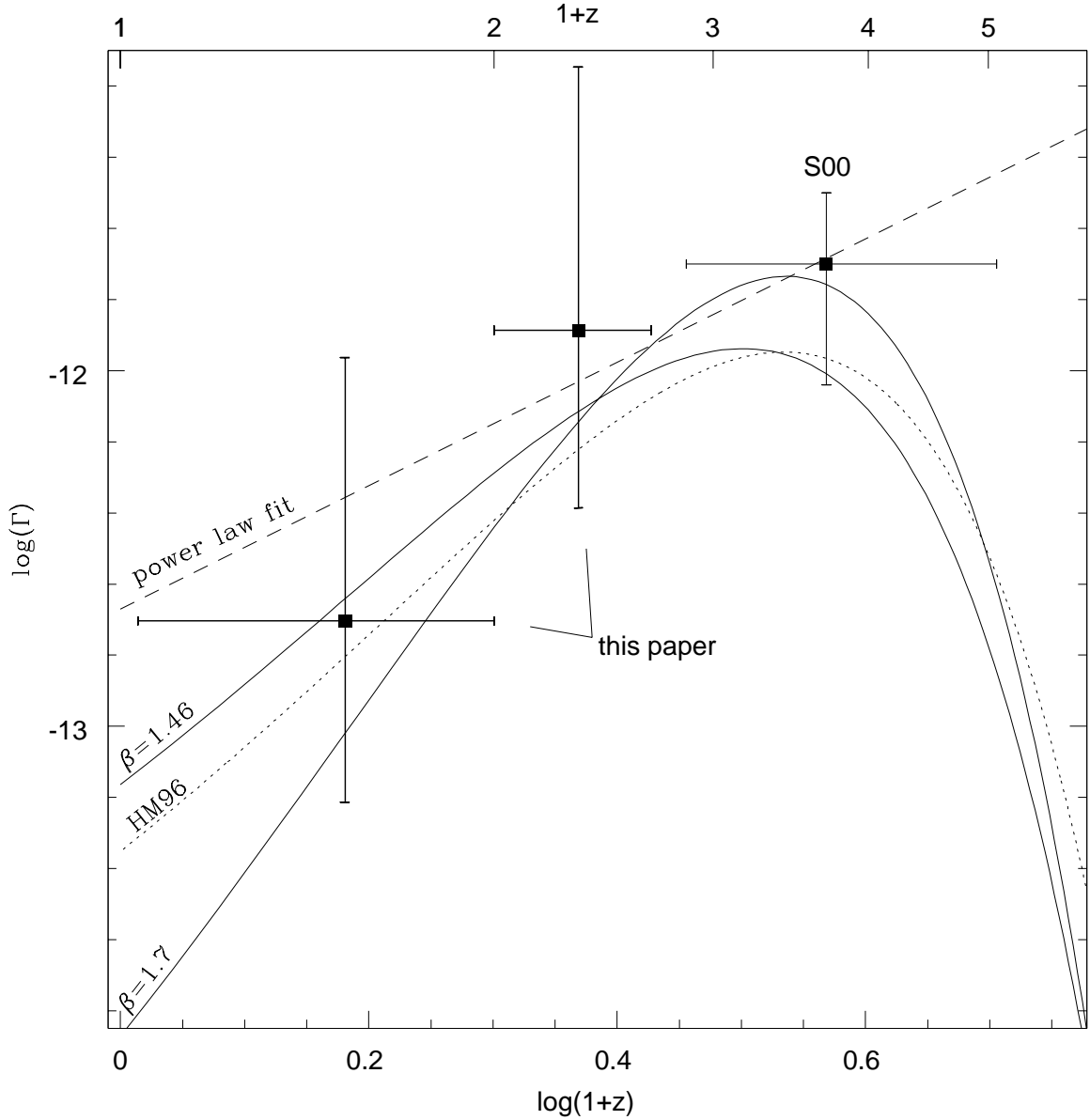}
\caption{HI ionization rate versus redshift:
(points)- constant equivalent width threshold maximum likelihood solutions
from this paper, at $z < 1$ and $z > 1$, and from Paper II for
$1.7 < z < 3.8$;
(dashed line)- constant threshold solution to Equ.~\ref{equ:plgam} for HST/FOS
data alone;
(solid line)- constant threshold solution to Equ.~\ref{equ:hmgam} with $\beta=1.46$
and $\beta=1.7$ for HST/FOS
data and ground-based data from Papers I and II,
(dotted line)- HM96 solution to Equ.~\ref{equ:hmgam}
\label{fig:gam} }
\end{figure}

\clearpage

\begin{figure}
\plotone{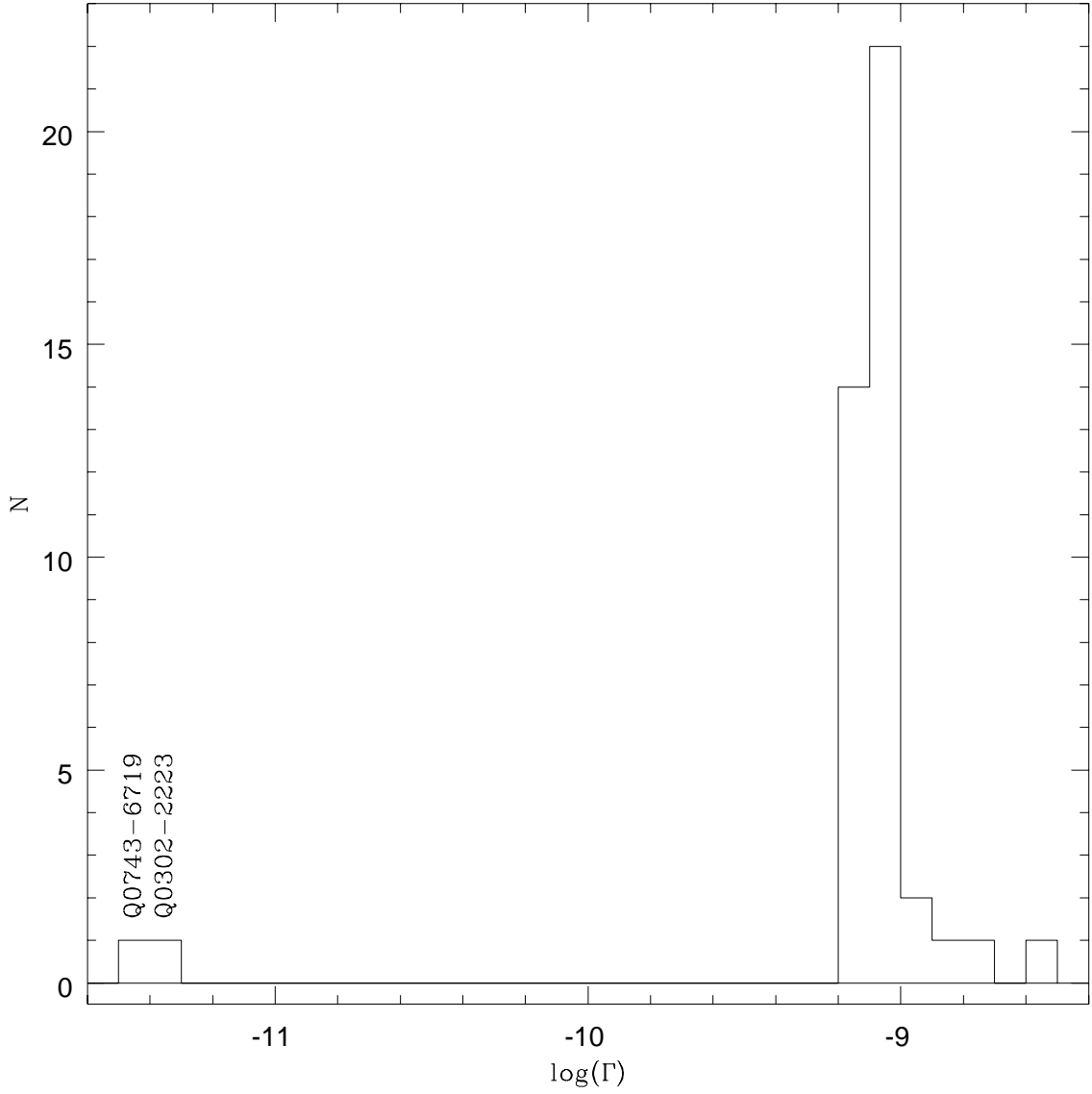}
\caption{Histogram of results of jackknife measurements of
HI ionization rate,
$\Gamma$, for all lines at $z > 1$ above
variable equivalent width threshold; labels on highest $\Gamma$ bins
indicate objects removed, see \S~\ref{sec-varthr}
\label{fig:jack} }
\end{figure}

\clearpage

\begin{figure}
\plotone{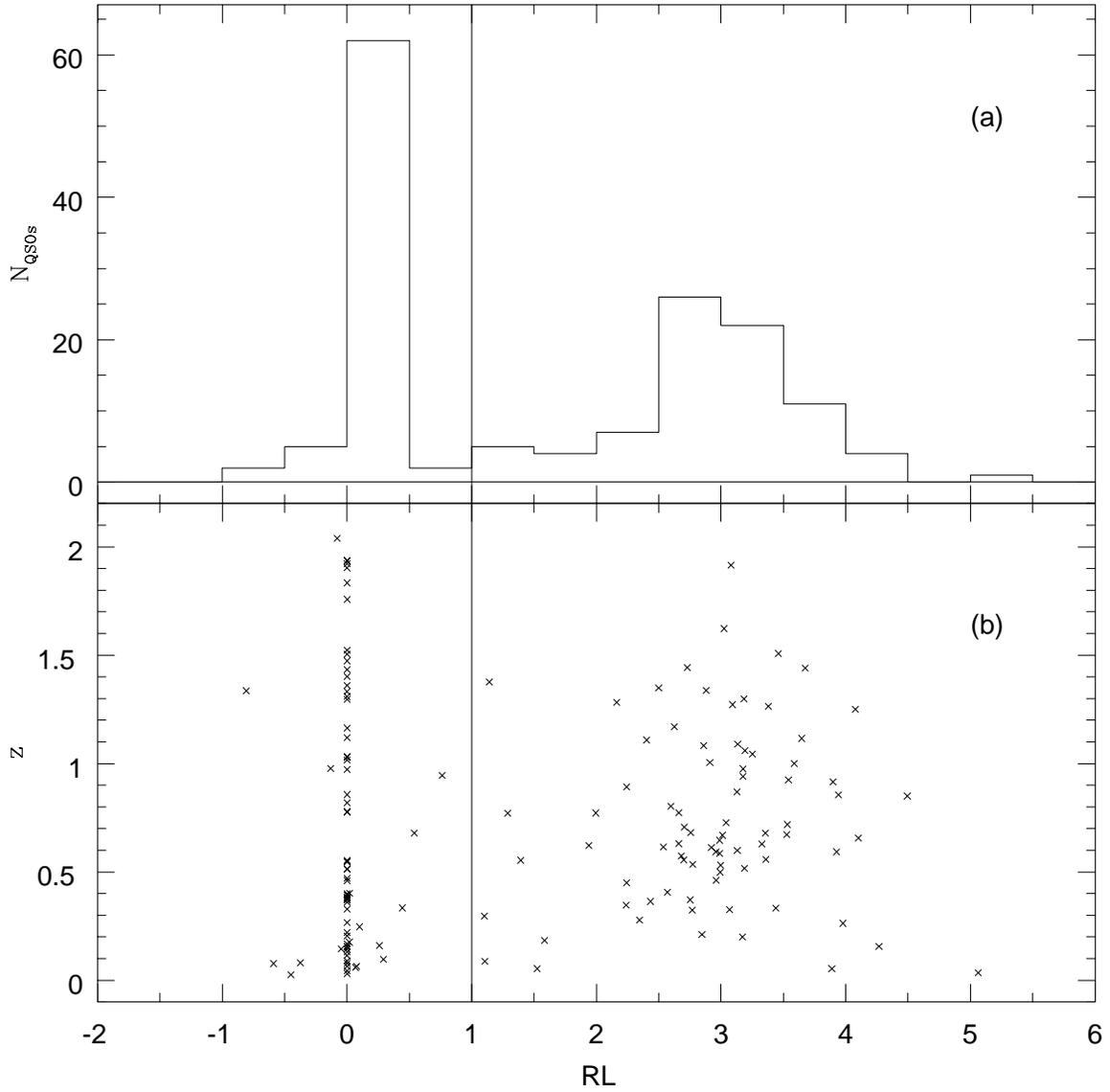}
\caption
{(a) Histogram of radio loudness (RL) values for QSOs in proximity effect sample,
where ${\rm RL =  log[S(5 \; GHz)]/log[S(1450} \; \mbox{\AA})]$,
includes blazars and objects with damped Ly-$\alpha$ absorption;
(b) redshift versus RL for QSOs in proximity effect sample
\label{fig:rl}}
\end{figure}

\clearpage

\begin{figure}
\plotone{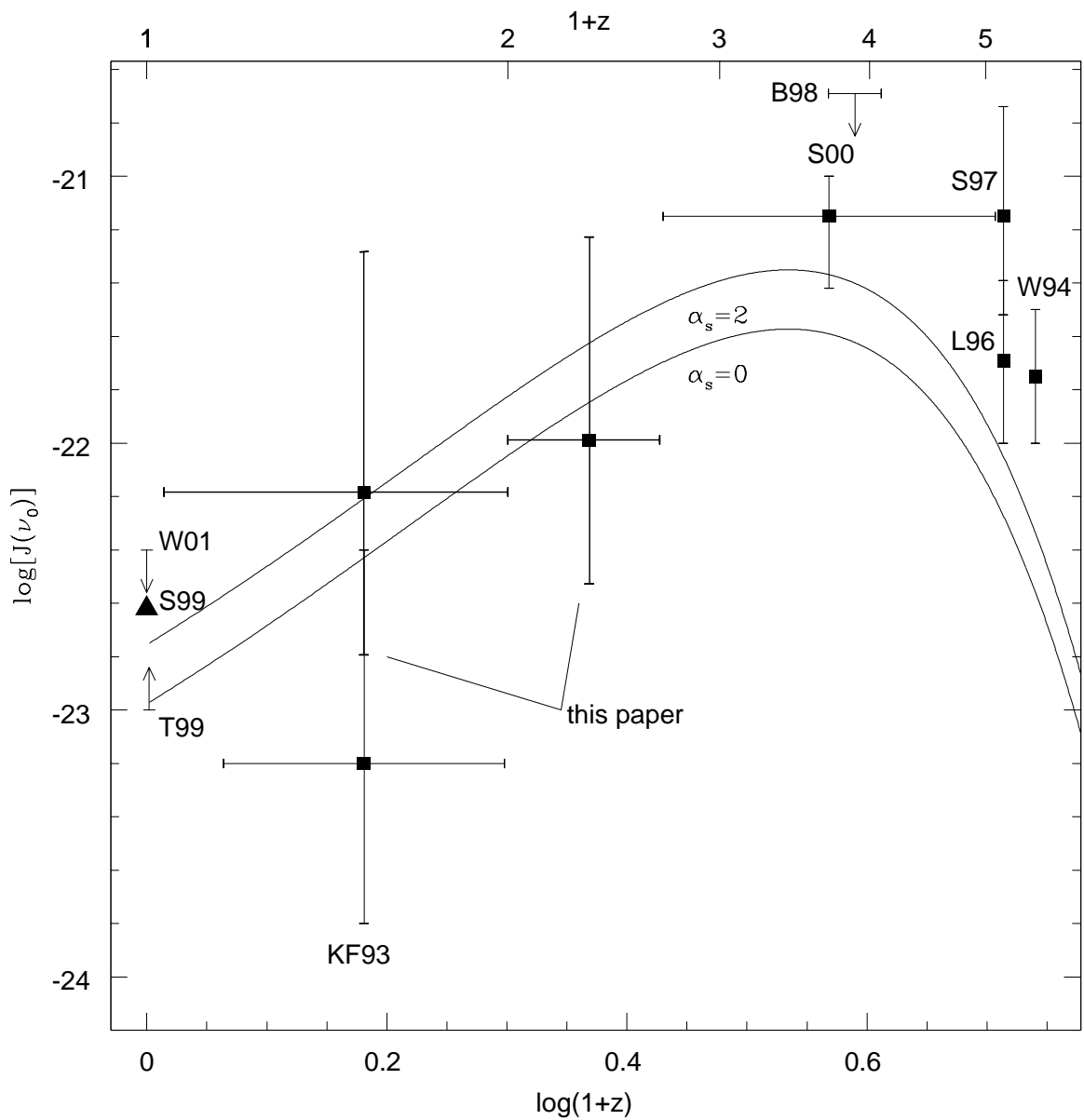}
\caption{log[$J(\nu_{0})$] versus redshift: 
(lower limit at $z \sim 0$)- Tumlinson et al.\ (1999);
(upper limit at $z \sim 0$)- Weymann et al.\ (2001); 
(filled triangle)- Shull et al.\ (1999);
(upper limit at $z=0$)- Weymann et al.\ (2001);
(filled squares, bold error bars)- our results for $z < 1$ and $z > 1$; 
(other filled squares)- results from
KF93, Paper II, Lu et al.\ (1996), Savaglio et al. (1997), 
and Williger et al.\ (1994);
(upper limit at $z \sim 3$)- Bunker et al.\ (1998);
(solid curves)- HM96 models for two values of the global source spectral
index, $\alpha_{s}$
\label{fig:allzcomp}}
\end{figure}

\clearpage

\begin{figure}
\plotone{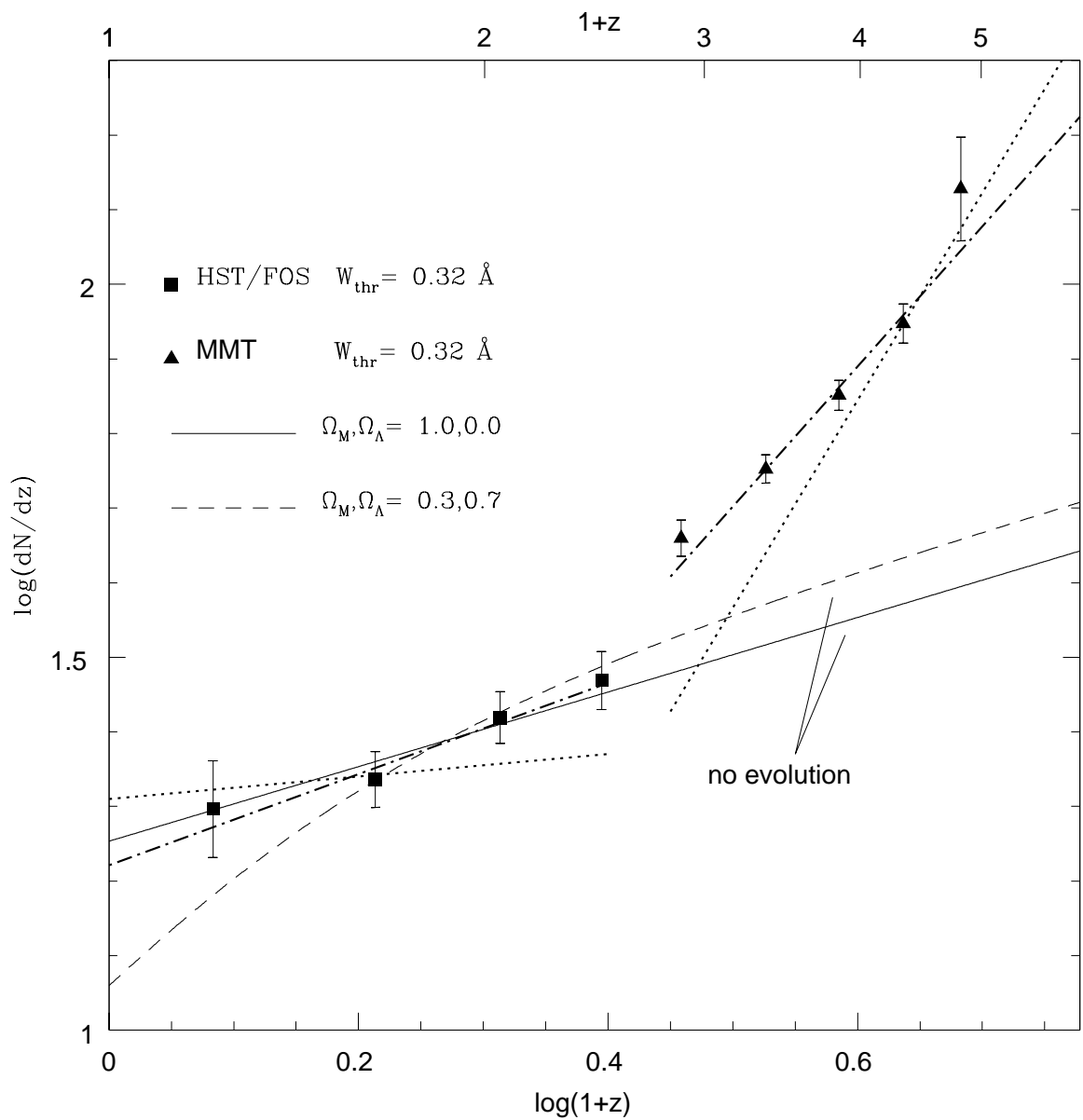}
\caption{$d{\cal N}/dz$ versus $z$:
solid and dashed lines show the relation for non-evolving Ly-$\alpha$ absorbers given by 
Equ.~\ref{equ:noevol} for $(\Omega_{M},\Omega_{\Lambda})=
(1.0,0.0)$ and $(0.3,0.7)$, respectively; dotted lines
are fits to low redshift data from Weymann et al.\ (1998) and to
high redshift data of Kim et al.\ (1997); dashed-dotted lines are fits to low
redshift data from Paper IV 
and to high redshift data from Paper I
\label{fig:noevol}}
\end{figure}

\clearpage

\begin{figure}
\plotone{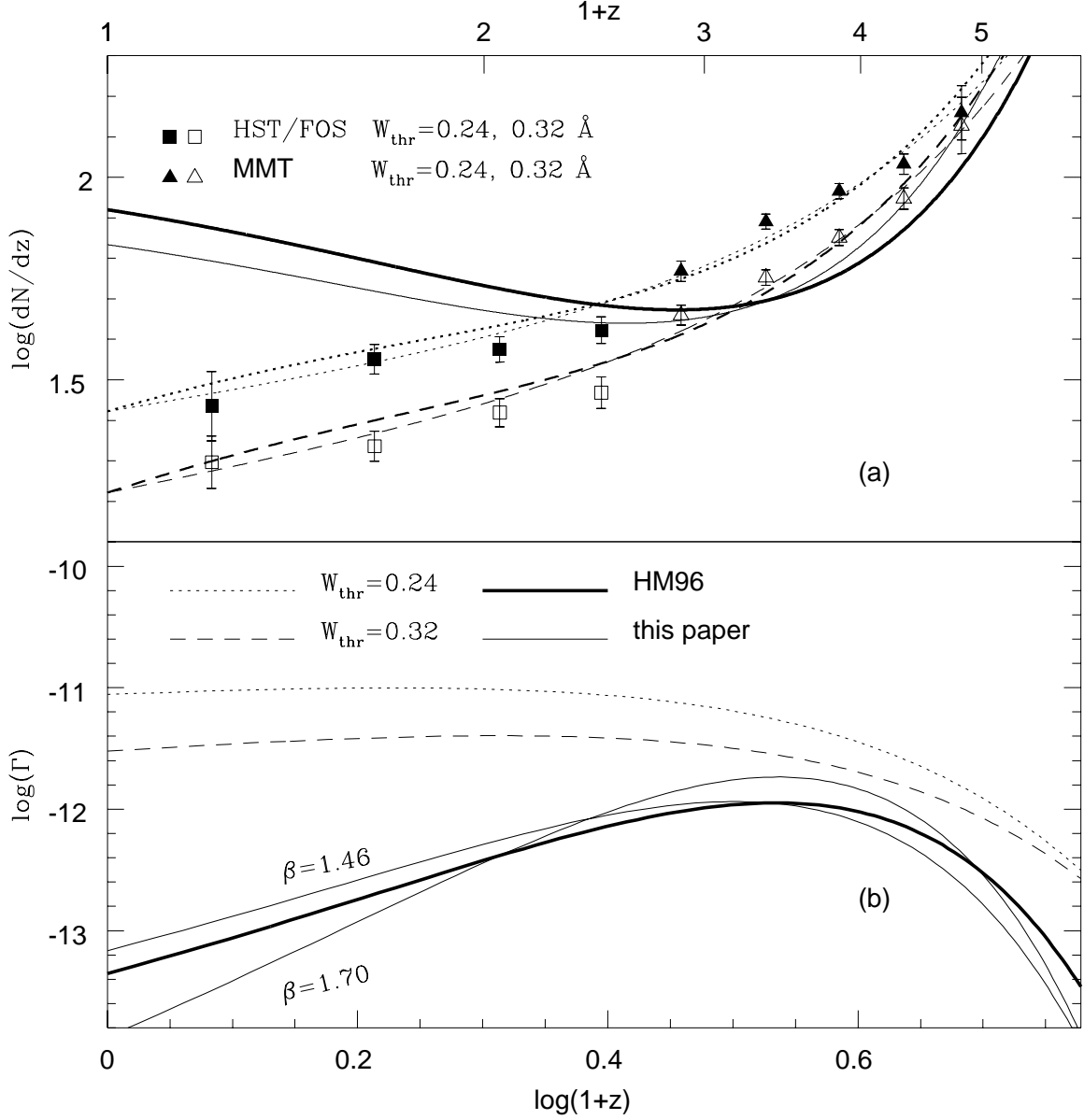}
\caption{(a) $d{\cal N}/dz$ versus $z$:
(solid points, dotted lines)
$W_{thr}=0.24$~\AA\ with fit to Equ.~\ref{equ:dave}, 
(open points, dashed lines)
$W_{thr}=0.32$~\AA\ with fit to Equ.~\ref{equ:dave},
(thick solid line)
Equ.~\ref{equ:dave} evaluated with HM96 parameters for 
$\Gamma(z)$ expressed by Equ.~\ref{equ:hmgam},
(thin solid lines)
Equ.~\ref{equ:dave} evaluated with parameters for $\Gamma(z)$
found in this paper; 
(b) $\Gamma(z)$ versus redshift expressed by
Equ.~\ref{equ:hmgam} using HM96 parameters (thick solid line), 
using parameters found in this paper (thin solid lines),
and using parameters found from 
fits to $d{\cal N}/dz$ for 
$W_{thr}=0.24$~\AA\ and $(\Omega_{M},\Omega_{\Lambda})=(1.0,0.0)$ (dotted line) and
$W_{thr}=0.32$~\AA\ and $(\Omega_{M},\Omega_{\Lambda})=(1.0,0.0)$ (dashed line),
\label{fig:dndz}}
\end{figure}

\clearpage

\begin{deluxetable}{llcccccccc}
\rotate
\tablecolumns{10}
\tablewidth{43pc}
\tablecaption{Sample QSOs and Emission Line Redshifts \label{tab-z}}
\tablehead{
\colhead{QSO\tablenotemark{1}} &\colhead{NED description}
&\colhead{Ly-$\alpha$} &\colhead{MgII}&\colhead{OIII} &\colhead{Balmer}
&\multicolumn{4}{c}{\underline{\hspace{0.5in}References\hspace{0.5in}}}\\
\colhead{} &\colhead{} &\colhead{(a)} &\colhead{(b)} &\colhead{(c)} &\colhead{(d)}
&\colhead{(a)} & \colhead{(b)} & \colhead{(c)} & \colhead{(d)} }
\startdata
0003+1553               &opt.var. &0.4497 &0.4502  &0.4503  &\nodata &(1) &(2) &(3) &    \\
0003+1955               &opt.var. &0.0264 &0.0264  &0.0261  &\nodata &(1) &(1) &(4) &    \\
0007+1041               &opt.var. &0.0902 &0.0890  &0.089   &0.0895  &(1) &(1) &(5) &(6) \\
0015+1612               &RQQ &0.5492 &\nodata &\nodata &\nodata &(1) &    &    &    \\
0017+0209               &LINER &0.3994 &\nodata &\nodata &\nodata &(1) &    &    &    \\
0024+2225               &\nodata &1.1081 &1.1096  &\nodata &\nodata &(1) &(7)&    &    \\
0026+1259               &Sy1 &0.1453 &0.1463  &0.1452  &0.1458  &(1) &(1) &(5) &(6) \\
0042+1010               &\nodata &0.5854 &0.583   &0.586   &0.584   &(1) &(8) &(8) &(8) \\
\enddata
\end{deluxetable}

\clearpage

\begin{deluxetable}{lcccc}
\tablecolumns{5}
\tablewidth{30pc}
\tablecaption{Summary of Observations \label{tab-zhst} }
\tablehead{
\colhead{Name}  &\colhead{V} &\colhead{Setup$^{\rm a}$} &
\colhead{Date}   &\colhead{Total exposure time (sec)} }
\startdata
0112-0142   &18.0 &1  &13Dec1996 &1200  \\
0137+0116   &17.1 &1  &13Dec1996 &1200  \\
0232-0415   &16.4 &1  &13Dec1996 &1200  \\
0349-1438   &16.2 &1  &12Dec1996 &900  \\
0414-0601   &15.9 &1  &19Dec1995 &400  \\ 
0454-2203   &16.1 &1  &19Dec1995 &400  \\
0624+6907   &14.2 &1  &19Dec1995 &465 \\
0827+2421   &17.2 &3  &15Feb1997 &1200  \\
0850+4400   &16.4 &1  &19Dec1995 &300 \\
0859-1403   &16.6 &2a &12Dec1996 &3600 \\
0916+5118   &16.5 &1      &19Dec1995 &350 \\
0923+3915   &17.9 &2b &14Jan1996 &1800 \\
0954+5537   &17.7 &2c &20Apr1996 &3600 \\
0959+6827   &16.4 &2b &14Jan1996 &1800 \\
1001+2910   &15.5 &2a &12Dec1996 &3600 \\
1008+1319   &16.2 &2a &10Dec1996 &1800 \\
1130+1108   &16.9 &2d &14Jan1996 &3600 \\
1138+0204   &17.6 &2e &12Dec1996 &2400  \\
1156+2123   &17.5 &2e &12Dec1996 &1800  \\ 
1156+2931   &17.0 &2a &10Dec1996 &1800  \\
1214+1804   &17.5 &2f &21Apr1996 &1800  \\
1230+0947   &16.1 &2f &21Apr1996 &3600 \\
1305+0658   &17.0 &2c &20Apr1996 &3600 \\
\tablenotetext{a}{Telescope/Instrument set up:  
(1) FLWO 1.5 m, FAST, 300~l~mm$^{-1}$ 1$^{\rm st}$ order, 3 $\arcsec$ slit, 3660-7540~\AA;
(2) SO B90, B\&C, 600~l~mm$^{-1}$ 1$^{\rm st}$ order, 1.5 $\arcsec$ slit, 
[a] 4500-6700~\AA, [b] 3600-5825~\AA, [c] 4140-6370~\AA,
[d] 6870-9140~\AA, [e] 5610-7860~\AA, [f] 5280-7550~\AA;
(3) MMT, Blue Channel, 800~l~mm$^{-1}$ 1$^{\rm st}$ order, 
2$\arcsec$ slit, 4365-6665~\AA}
\enddata 
\end{deluxetable}

\clearpage

\begin{deluxetable}{lcccccrcccc}
\tablewidth{42pc}
\tablecaption{Spectrophotometric Properties \label{tab-flux}}
\tablecolumns{11}
\tablehead{
\colhead{QSO} &\colhead{N$_{HI}$} &\colhead{f$_{\nu_{0}}^{\rm obs}$} & \colhead{$\alpha$}
&\colhead{f$_{\nu_{0}}$} &\colhead{f$_{\nu}^{obs}$}& \colhead{RL} 
&\multicolumn{4}{c}{\underline{\hspace{0.3in}References\hspace{0.3in}}}\\
\colhead{} & \colhead{(a)} & \colhead{(b)} & \colhead{(c)} &\colhead{(d)} &\colhead{(e)} &\colhead{(f)}  
& \colhead{(b)} & \colhead{(c)} &\colhead{(d)} &\colhead{(e)} }
\startdata
0003+1553                 &3.88 &0.46 & 0.71$\pm$0.52 &1.39$\pm$0.33 &1.94 (1450)& 2.24 &(2)&(1b)&(1b)&(1b)\\
0003+1955                 &3.99 &2.04 & 0.47$\pm$0.09 &6.77$\pm$0.45 &8.43 (1450)&-0.44 &(3)&(1a)&(1a)&(1a)\\
0007+1041                 &5.62 &     &-0.50$\pm$1.00 &1.86$\pm$0.66 &1.47 (1450)& 0.00 &   &(1a)&(1a)&(1a)\\
0015+1612                 &4.07 &     &-1.14$\pm$0.43 &0.19$\pm$0.06 &0.11 (1450)& 0.00 &   &(1b)&(1b)&(1b)\\
0017+0209                 &3.05 &     & 1.98$\pm$0.56 &0.12$\pm$0.08 &0.31 (1450)& 0.00 &   &(1b)&(1b)&(1b)\\
0024+2225\tablenotemark{1}&3.60 &     & 0.59$\pm$0.65 &0.60$\pm$0.15 &0.79 (1450)& 2.40 &   &(1c)&(1c)&(1c)\\ 
0026+1259                 &4.56 &     &-0.10$\pm$0.24 &2.33$\pm$0.33 &2.22 (1450)&-0.04 &   &(1b)&(1b)&(1b)\\
0042+1010                 &5.52 &     & 0.19$\pm$0.08 &0.08$\pm$0.02 &0.09 (1450)& 2.99 &   &(1c)&(1c)&(1c)\\
\enddata
\tablecomments{ Columns
(a) 10$^{20}$ cm$^{-2}$ from 
Stark et al.\ 1992; Burstein \& Heiles 1982; Lockman \& Dickey 1995; 
(b) Direct measurement of flux at Lyman Limit in $\mu$Jy;
(c) Measured spectral index;
(d) Extrapolated flux at Lyman Limit in $\mu$Jy;
(e) Measured flux at the rest wavelength listed in \AA;
(f) Radio Loudness, ${\rm RL =  log[S(5 \; GHz)]/log[S(1450} \; \AA )]$ }
\tablerefs{ 
(1) this paper, FOS data [a] H130, [b] H190, [c] H270; 
(2)Lanzetta et al.\ 1993;
(3)Zheng et al.\ 1995;
(4)Zheng et al.\ 1997;
(5)Hamann et al.\ 1995;
(6) Tytler \& Fan 1992;
(7)Appenzeller et al.\ 1998;
(8)Sanduleak \& Pesch 1984;
(9)Osmer et al.\ 1994;
(10) Zheng \& Malkan 1993;
(11)Kinney et al.\ 1991;
(12) Green 1996 \\}
\tablenotetext{1}{Object spectrum shows associated absorption}
\tablenotetext{2}{Observed only with pre-COSTAR FOS and A-1 aperture, not used for proximity effect}
\tablenotetext{3}{Object is classified as a blazar or BL Lac }
\tablenotetext{4}{Flux estimated from scaling composite QSO spectrum to match V}
\tablenotetext{5}{Object spectrum shows damped Ly-$\alpha$ absorption}
\tablenotetext{6}{No Ly-$\alpha$ forest observed, not used for proximity effect}
\tablenotetext{7}{Binary quasar, not used for proximity effect}
\tablenotetext{8}{Flux estimated from scaling composite QSO spectrum to match B}
\end{deluxetable}

\clearpage

\begin{deluxetable}{llclcccccc}
\tablecolumns{10}
\tablewidth{40pc}
\tablecaption{Measurements of $J(\nu_{0})$ 
\label{table-jnu} } 
\tablehead{\colhead{Sample} &\colhead{${\cal N}_{lines}$} & 
\colhead{$\gamma$,norm.}  & \colhead{$\beta$} & \colhead{b} & 
\colhead{method} & \colhead{log[$J(\nu_{0})$]}  & \colhead{$\chi^{2}$} &
\colhead{Q$_{\chi^{2}}$} &\colhead{Q$_{\rm KS}$} \\
\colhead{(a)}  &\colhead{(b)} &\colhead{(c)} &\colhead{} &\colhead{(d)} &
\colhead{(e)} &\colhead{(f)} &\colhead{(g)} &\colhead{(h)} &
\colhead{(i)} }
\startdata
1 &259  &0.82, 13.6 &1.46 &\nodata &BDO &-22.04$^{+0.43}_{-1.11}$  &2.13 &0.95 &0.80 \\
1 &259  &0.82, 6.73 &1.46 &35 &ML &-22.11$^{+0.51}_{-0.40}$ &1.21 &0.29 &0.80 \\
1 &259  &0.82, 9.61 &1.46 &25 &ML &-22.12$^{+0.52}_{-0.39}$ &1.01 &0.41 &0.80 \\ 
1 &259  &0.82, 9.31 &1.45 &25 &ML &-22.13$^{+0.51}_{-0.41}$ &0.78 &0.58 &0.80 \\
1 &259  &0.82, 11.8 &1.70 &30 &ML &-21.74$^{+0.45}_{-0.36}$ &1.34 &0.23 &0.80 \\
1 &259  &0.82, 38.0 &2.04 &25 &ML &-21.47$^{+0.43}_{-0.32}$ &1.10 &0.35 &0.80 \\
2 &289  &0.15, 31.3 &1.46 &\nodata &BDO &-22.06$^{+0.05}_{-0.62}$ &2.62 &0.91 &0.30 \\
2 &289  &0.15, 12.0 &1.46 &35 &ML &-22.03$^{+0.44}_{-0.37}$ &1.32 &0.24 &0.30 \\
\enddata
\end{deluxetable}

\clearpage

\begin{deluxetable}{cclccc}
\tablecolumns{6}
\tablewidth{39pc}
\tablecaption{Simulation Results \label{table-sim}} 
\tablehead{
\colhead{Input log[$J(\nu_{0})$]} &\colhead{$z$} &\colhead{$W_{thr}$} &\colhead{$\gamma$,$A$} &
\colhead{Recovered log[$J(\nu_{0})$]} &\colhead{$\chi^{2}$}  \\
\colhead{(a)}  & \colhead{(b)} & \colhead{(c)} &\colhead{(d)} 
&\colhead{(e)} & \colhead{(f)} }
\startdata
-23.0 &all   &0.32     &1.41,7.81 &-22.74$^{+0.15}_{-0.13}$ &5.13 \\
-23.0 &all   &variable &1.15,8.94 &-22.47$^{+0.13}_{-0.12}$ &5.73 \\
-22.0 &all   &0.32     &1.17,8.27 &-21.32$^{+0.41}_{-0.29}$ &13.6 \\
-22.0 &$z<1$ &0.32     &0.95,8.74 &-20.81$^{+1.91}_{-0.62}$ &9.53 \\
-22.0 &$z>1$ &0.32     &1.79,5.21 &-21.64$^{+0.39}_{-0.28}$ &6.70 \\ 
-22.0 &all   &variable &1.48,6.71 &-21.63$^{+0.20}_{-0.19}$ &3.21 \\
-22.0 &$z<1$ &variable &0.75,10.0 &-21.34$^{+0.60}_{-0.36}$ &11.3 \\
-22.0 &$z>1$ &variable &1.52,6.09 &-21.63$^{+0.25}_{-0.22}$ &1.30 \\
\enddata
\end{deluxetable}

\clearpage

\begin{deluxetable}{llcccccc}
\tablecolumns{8}
\tablewidth{33pc}
\tablecaption{HI Ionization Rates \label{table-gamma}} 
\tablehead{
\colhead{Sample} &\colhead{$\gamma$,$A$} &\colhead{$\beta$} &\colhead{b}
&\colhead{log[$\Gamma_{\rm HI}$]} &\colhead{$\chi_{\nu}^{2}$} &\colhead{$Q_{\chi_{\nu}^{2}}$} &\colhead{log[$J(\nu_{0})$]} \\
\colhead{(a)}  &\colhead{(b)} &\colhead{} &\colhead{} &\colhead{} 
&\colhead{(c)} &\colhead{(d)} &\colhead{(e)}}
\startdata
1  &0.69,7.65 &1.46 &35 &-12.17$^{+0.50}_{-0.40}$ &0.49 &0.81 &-21.56 \\
1a &0.85,7.11 &1.46 &35 &-12.70$^{+0.74}_{-0.51}$ &0.38 &0.88 &-22.09 \\
1b &0.72,7.29 &1.46 &35 &-11.88$^{+0.74}_{-0.50}$ &0.48 &0.81 &-21.28 \\
2  &0.61,9.27 &1.46 &35 &-11.27$^{+0.74}_{-0.45}$ &0.78 &0.58 &-20.67 \\
2a &0.63,9.24 &1.46 &35 &-12.23$^{+0.55}_{-0.42}$ &1.17 &0.31 &-21.62 \\
2b &1.05,6.40 &1.46 &35 &-9.089$^{+\infty}_{-2.22}$ &1.17 &0.31 &-18.48 \\
1  &0.69, 7.21\tablenotemark{1} &1.46 &35 &-12.67,1.73\tablenotemark{1} &1.01 &0.40 &\nodata \\
2  &0.61, 9.04\tablenotemark{1} &1.46 &35 &-10.86,3.04\tablenotemark{1} &0.47 &0.82 &\nodata \\
\tablecomments{ (a) (1) all lines with $W>0.32$~\AA, (1a) $z<1$, (1b) $z>1$,
(2)all lines above variable threshold, (2a) $z<1$, (2b) $z>1$;
(b) Maximum likelihood method normalization
(see \S\ref{sec-results}, Equ.~\ref{eq:mlnorm});
(c) $\chi^{2}$ of data versus the ionization model used;
(d) $\chi^{2}$ probability for the ionization model used;
(e) $J(\nu_{0})$ implied by $\Gamma$ listed and $\alpha_{s}=1.8$ 
(see \S\ref{sec-gam}, Equ.~\ref{eq:gamma}) }
\tablenotetext{1}{maximum likelihood solution for $\log(A_{\rm pl})$,$B_{\rm pl}$ and normalization 
(see \S\ref{sec-gam}, Equ.~\ref{equ:plgam})}
\enddata
\end{deluxetable}

\end{document}